\documentclass[twocolumn]{aastex61}

\shorttitle{Probing the ionization regions of the stellar convective envelopes}
\shortauthors{Brito \& Lopes}

\begin{document}

\title{From cool to hot F-stars: the influence of two ionization regions in the acoustic oscillations}

\correspondingauthor{Ana Brito}
\email{ana.brito@tecnico.ulisboa.pt}
\email{ilidio.lopes@tecnico.ulisboa.pt}

\author{Ana Brito}
\affil{Centro de Astrof\'isica e Gravita\c{c}\~ao\\ Instituto Superior T\'ecnico, Universidade de Lisboa \\ Av. Rovisco Pais, 1049-001 Lisboa, Portugal}
\affil{Departamento de Matem\'atica, Instituto Superior de Gest\~ao\\ Av. Marechal Craveiro Lopes, 1700-284, Lisboa, Portugal}

\author{Il\'idio Lopes}
\affiliation{Centro de Astrof\'isica e Gravita\c{c}\~ao\\ Instituto Superior T\'ecnico, Universidade de Lisboa \\ Av. Rovisco Pais, 1049-001 Lisboa, Portugal}

\begin{abstract}

  The high-precision data available from the \textit{Kepler} satellite allows us to venture in the study of the complex outer convective envelopes of solar-type stars. We use a seismic diagnostic, specialized for investigating the outer layers of  solar-type stars, to infer the impact of the ionization processes on the oscillation spectrum, for a sample of \textit{Kepler} stars. These stars, of spectral type F, cover all of the observational seismic domain of the acoustic oscillation spectrum in solar-type stars. They also cover the range between a cool F-dwarf ($\sim 6000$ K)  and a hotter F-star ($\sim 6400$ K).
 
 Our study reveals the existence of two relevant ionization regions. One of these regions, which is located closer to the surface of the star, is commonly associated with the second ionization of helium, although other chemical species also contribute to ionization. The second region, located deeper in the envelope, is linked with the ionization of heavy elements. Specifically, in this study, we analyze the elements carbon, nitrogen, oxygen, neon, and iron. Both regions can be related to the K electronic shell. We show that, while for cooler stars like the Sun, the influence of this second region on the oscillation frequencies is small; in hotter stars, its influence becomes comparable to the influence of the region of the second ionization of helium. This can guide us in the study of the outer layers of F-stars, specifically with the understanding of phenomena related to rotation and magnetic activity in these stars.

\end{abstract}


\section{Introduction} \label{sec:intro}

Asteroseismology gives us the possibility to probe the stellar interior. This is done through the interpretation of the oscillations of stars \citep[e.g.,][]{2013ARA&A..51..353C}, which are utterly determined by the properties of the stellar structure. The space missions {\it{CoRoT}} \citep{2007AIPC..895..201B, 2008CoAst.156...73M} and {\it{Kepler}} \citep{2010Sci...327..977B, 2010PASP..122..131G} have dramatically improved the observational datasets, which provide the base to test and improve, the theories of stellar structure and evolution.
Solar-type stars, such as main-sequence F-stars, are among the targets of these space missions. They exhibit several different patterns of magnetic activity \citep[e.g.,][]{2014A&A...562A.124M, 2017A&A...598A..77K}. The surface magnetism, which can also be traced by photometric methods, is correlated to the rotation rates in these stars \citep[e.g.,][]{2014A&A...572A..34G, 2015SSRv..196..303B, 2015sac..book..437C}. The exact origin of this correlation, as well as the origins of the stellar dynamos presumably acting in these stars, is still unclear \citep[e.g.,][]{2016ApJ...826L...2M, 2016Natur.535..526W, 2017arXiv170102591A}.

The outer layers of F-stars are of major interest when investigating all the phenomena related with magnetic activity. After all, these upper layers are the stage of a complex and interconnected web of physical processes taking place at the same timescales, such as convection, differential rotation, and the generation of magnetic fields. For this same reason, the outer convective envelopes of solar-type stars are regions of great uncertainty. One of the major physical processes occurring in the outer layers of solar-type stars is partial ionization. A detailed study of observable data can help us to understand how the partial ionization of elements relates to rotation and magnetism in these stars.

%%%%%%%%%%%%%%%%%%%%%%%%%%%%%%%%%%%%%%%%%%%%%%%%%%%%%%%%%%%%%%%%%%%%%%%%%%%%%%%%%%%%%%%%
%%%%%  TABLE 1
%%%%%%%%%%%%%%%%%%%%%%%%%%%%%%%%%%%%%%%%%%%%%%%%%%%%%%%%%%%%%%%%%%%%%%%%%%%%%%%%%%%%%%%%
\begin{deluxetable*}{ccCccc} 
	\tablecaption{Observational constraints for the selected stars \tablenotemark{a)} \label{table:1}}
	\tablecolumns{6}
	\tablenum{1}
	\tablewidth{0pt}
	\tablehead{
		\colhead{Star Id.} &
		\colhead{$\langle \Delta \nu \rangle$ ($\mu $Hz) } &
		\colhead{$\nu_{\text{max}}$ ($\mu $Hz)} &
		\colhead{$T_{\text{eff}}$ (K) } &
		\colhead{[Fe/H] (dex)}  &
		\colhead{log $g$ } 
	}
	\startdata 
	A &$120.288^{+0.017}_{-0.018}$  &$2795.3^{+6.0}_{-5.7}$&$6067 \pm 120$&$-0.10 \pm 0.15$&$4.388^{+0.007}_{-0.008}$ \\ 
	B &$103.179^{+0.027}_{-0.027}$  &$2203.7^{+6.7}_{-6.3}$&$6140 \pm 77$ &$-0.19 \pm 0.10$&$4.277^{+0.011}_{-0.011}$ \\ 
	C &$70.369^{+0.034}_{-0.033}$ &$1406.7^{+6.3}_{-8.4}$&$6326 \pm 77$ &$-0.01 \pm 0.10$&$4.100^{+0.009}_{-0.009}$ \\ 
	D &$51.553^{+0.046}_{-0.047}$   &$958.3^{+4.6}_{-3.6}$ &$6331 \pm 77$ \tablenotemark{b)} &$-0.05 \pm 0.10$&$3.942^{+0.007}_{-0.005}$ \\
	\enddata
	\tablenotetext{a) }{  All values taken from \citet{2017ApJ...835..172L}}
	\tablenotetext{b) }{  In the modeling of the star D, we used the value for $T_{\text{eff}}$ from \citet{2014ApJS..210....1C} ($6499\pm46$ K)}
\end{deluxetable*}
%%%%%%%%%%%%%%%%%%%%%%%%%%%%%%%%%%%%%%%%%%%%%%%%%%%%%%%%%%%%%%%%%%%%%%%%%%%%%%%%%%%%%%%%
%%%%%  TABLE 1 - END
%%%%%%%%%%%%%%%%%%%%%%%%%%%%%%%%%%%%%%%%%%%%%%%%%%%%%%%%%%%%%%%%%%%%%%%%%%%%%%%%%%%%%%%%

%%%%%%%%%%%%%%%%%%%%%%%%%%%%%%%%%%%%%%%%%%%%%%%%%%%%%%%%%%%%%%%%%%%%%%%%%%%%%%%%%%%%%%%%
%%%%%  TABLE 2
%%%%%%%%%%%%%%%%%%%%%%%%%%%%%%%%%%%%%%%%%%%%%%%%%%%%%%%%%%%%%%%%%%%%%%%%%%%%%%%%%%%%%%%%
\begin{deluxetable*}{ccccccccc} 
	\tablecaption{Properties of the theoretical stellar models \label{table:2}}
	\tablecolumns{9}
	\tablenum{2}
	\tablewidth{0pt}
	\tablehead{
		\colhead{Star Id.} &
		\colhead{Age (Gyr)} &
		\colhead{${M}/{M}_\odot$} &
		\colhead{${R}/{R}_\odot$} & 
		\colhead{${L}/{L}_\odot$} & 
		\colhead{$T_{\text{eff}}$ (K)} &
		\colhead{$Y_0$} &
		\colhead{$Z$} &
		\colhead{$r_{\text{bcz}}/R$}
	}
	\startdata 
	A & 3.259 & 1.10 & 1.103 & 1.434 & 6019 & 0.264 & 0.0196 & 0.751 \\ 
	B & 6.459 & 1.05 & 1.210 & 1.864 & 6136 & 0.260 & 0.0150 & 0.732 \\ 
	C & 2.582 & 1.39 & 1.665 & 4.146 & 6388 & 0.265 & 0.0254 & 0.839 \\ 
	D & 2.782 & 1.39 & 2.033 & 6.546 & 6481 & 0.278 & 0.0201 & 0.881 \\ 
	\enddata
\end{deluxetable*}
%%%%%%%%%%%%%%%%%%%%%%%%%%%%%%%%%%%%%%%%%%%%%%%%%%%%%%%%%%%%%%%%%%%%%%%%%%%%%%%%%%%%%%%%
%%%%%  TABLE 2 - END
%%%%%%%%%%%%%%%%%%%%%%%%%%%%%%%%%%%%%%%%%%%%%%%%%%%%%%%%%%%%%%%%%%%%%%%%%%%%%%%%%%%%%%%%

In this work, we propose to use the method of the acoustic potential \citep{1989SvAL...15...27B, 1989ASPRv...7....1V, 1995ARep...39..105B, 1997ApJ...480..794L, 2001MNRAS.322..473L} to analyze the ionization processes occurring in the theoretical convective envelopes of solar-type stars. 
We chose four benchmark stars from the {\it{Kepler}} Legacy sample \citep{2017ApJ...835..172L} that cover all the range of observed large frequency separations. This means an interval from approximately $52 \, \mu$Hz (a hot F-star)  to $120 \, \mu$Hz (a cool dwarf). The stars of the Legacy sample are known to have the highest signal-to-noise ratios (S/N) among all of the {\it{Kepler}} stars. They have, in turn, the most precise set of oscillation mode frequencies.

Using the ionization fractions for seven chemical elements---hydrogen, helium, carbon, nitrogen, oxygen, neon, and iron---we study the variations of the mean ionic charges with depth and relate them to the acoustic potential and also to the seismic indicator $\beta(\nu)$, a proxy of the phase shift frequency dependence $\alpha(\nu)$, particularly sensitive to partial ionization processes in the outer layers of solar-type stars \citep{1987SvAL...13..179B, 1989SvAL...15...27B, 1989ASPRv...7....1V}. Ionization is about local changes in the number of particles. Our analysis exposes two relevant, ionization related regions, in the outer layers of solar-type stars. These regions can be associated with the ionizations of the electrons of the K-shell, reflecting the atomic periodic structure of matter. The closest to the surface of these two regions, which we named region K1 because it corresponds to the complete ionization of the first "period"(first line) in the periodic table of elements, is the expected region of the second ionization of helium. The second relevant ionization region lies deeper in the convection zone and is related to the ionization of the K-shell electrons of elements in the second "period" (second line), and was naturally named region K2.

\begin{figure} 
	\plotone{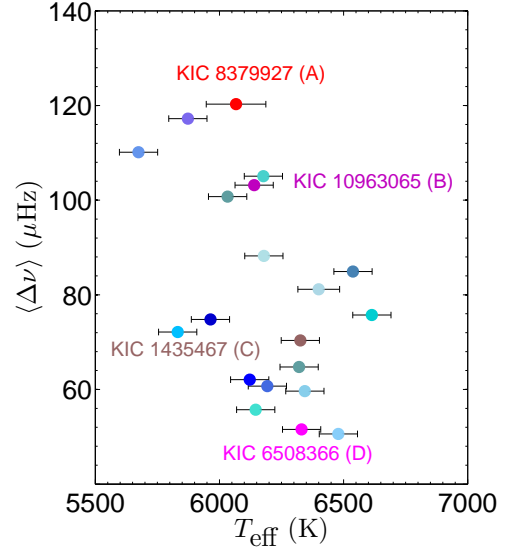}
	\caption{Asteroseismic HR diagram for 20 stars observed by {\it{Kepler}}. Error bars for the effective temperature are shown in black. Error bars for large separation values are too small to be distinguishable. The four selected solar-type stars of this study are identified with the corresponding KIC values in the figure and below with bold text.  All stars from top to bottom: {\bf{KIC 8379927 (A)}}, KIC 8760414, KIC 6603624, KIC 10454113, {\bf{KIC 10963065 (B)}}, KIC 6116048, KIC 12009504, KIC 9206432, KIC 9139163, KIC 2837475, KIC 1225814, KIC 6933899, {\bf{KIC 1435467 (C)}}, KIC 9812850, KIC 8228742, KIC 3632418, KIC 7103006, KIC 10162436, {\bf{KIC 6508366 (D)}}, and KIC 6679371. }
\end{figure}\label{fig1}

The paper is organized as follows. In Section \ref{sec2} we describe the theoretical models of the chosen benchmark stars and then we use the obtained theoretical models to characterize the different ionization patterns in the interiors of these stars in Section \ref{sec3}. The reflecting acoustic potentials and the corresponding theoretical seismic signatures $\beta(\nu)$ are discussed and related to the relevant ionization regions in Section \ref{sec4}. Section \ref{sec5} discusses the comparison between the observational and theoretical signatures $\beta(\nu)$. Finally, we present our conclusions in Section \ref{sec6}.

\begin{figure*}
	\plottwo{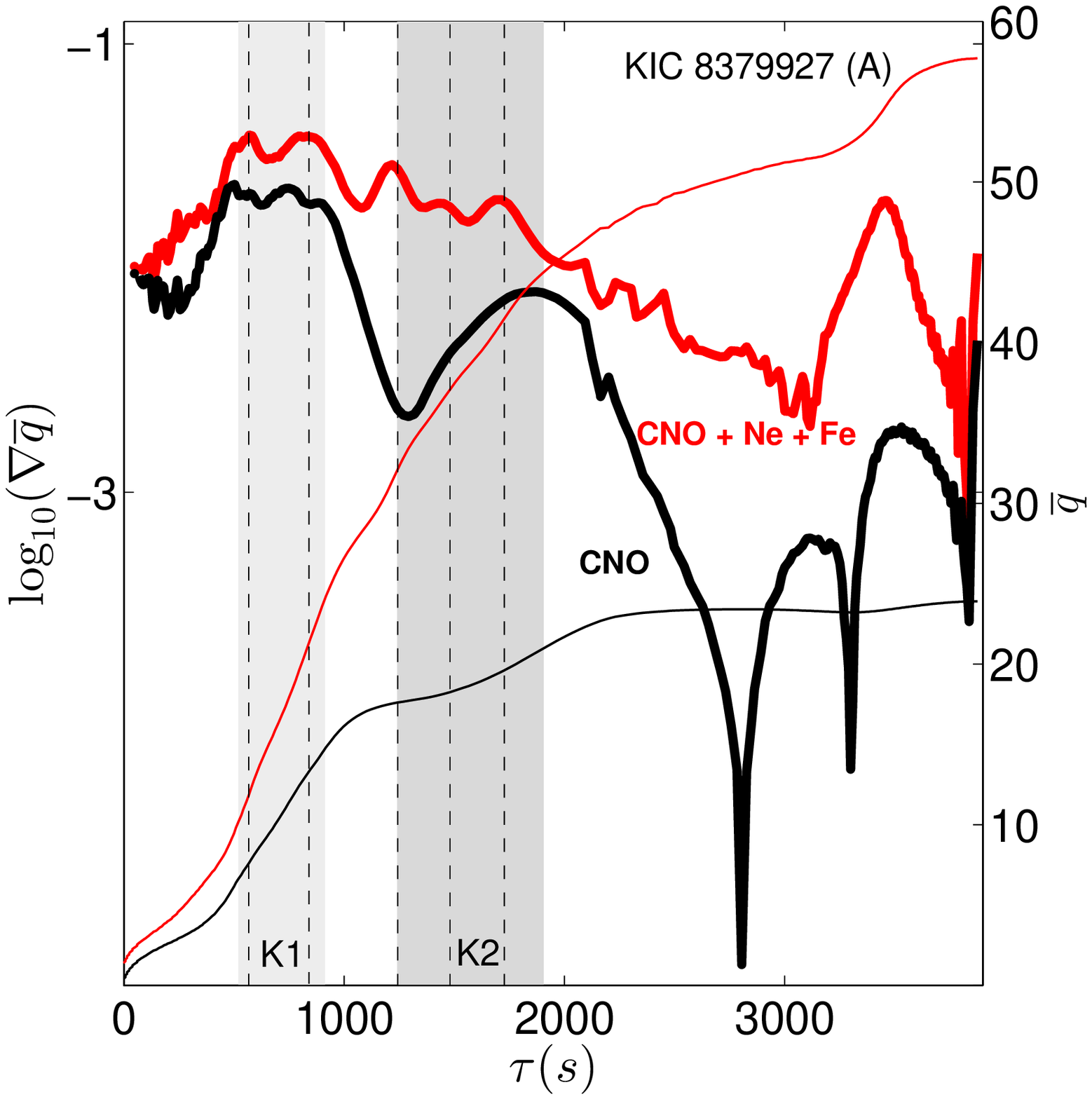}{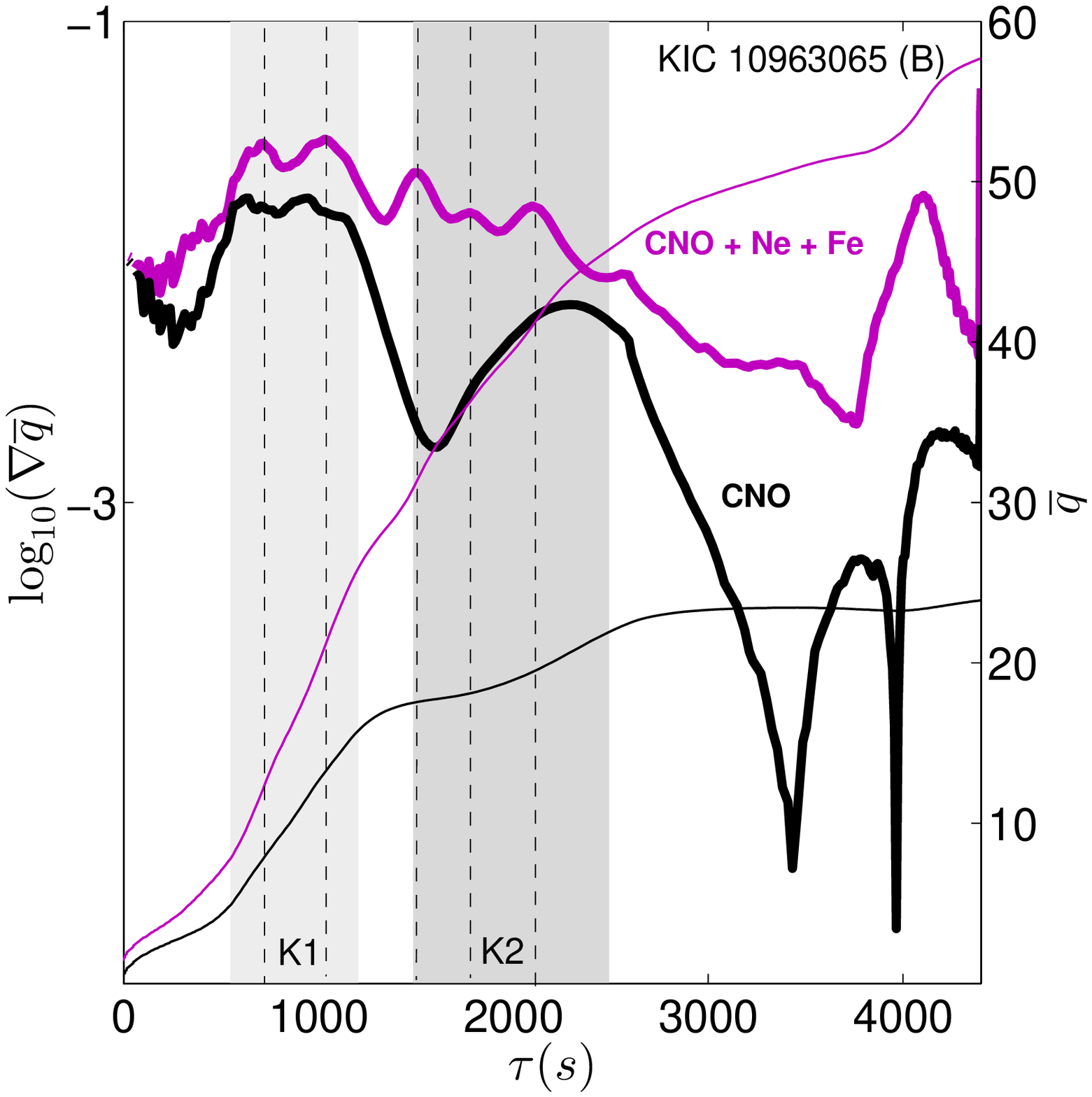}\\
	\plottwo{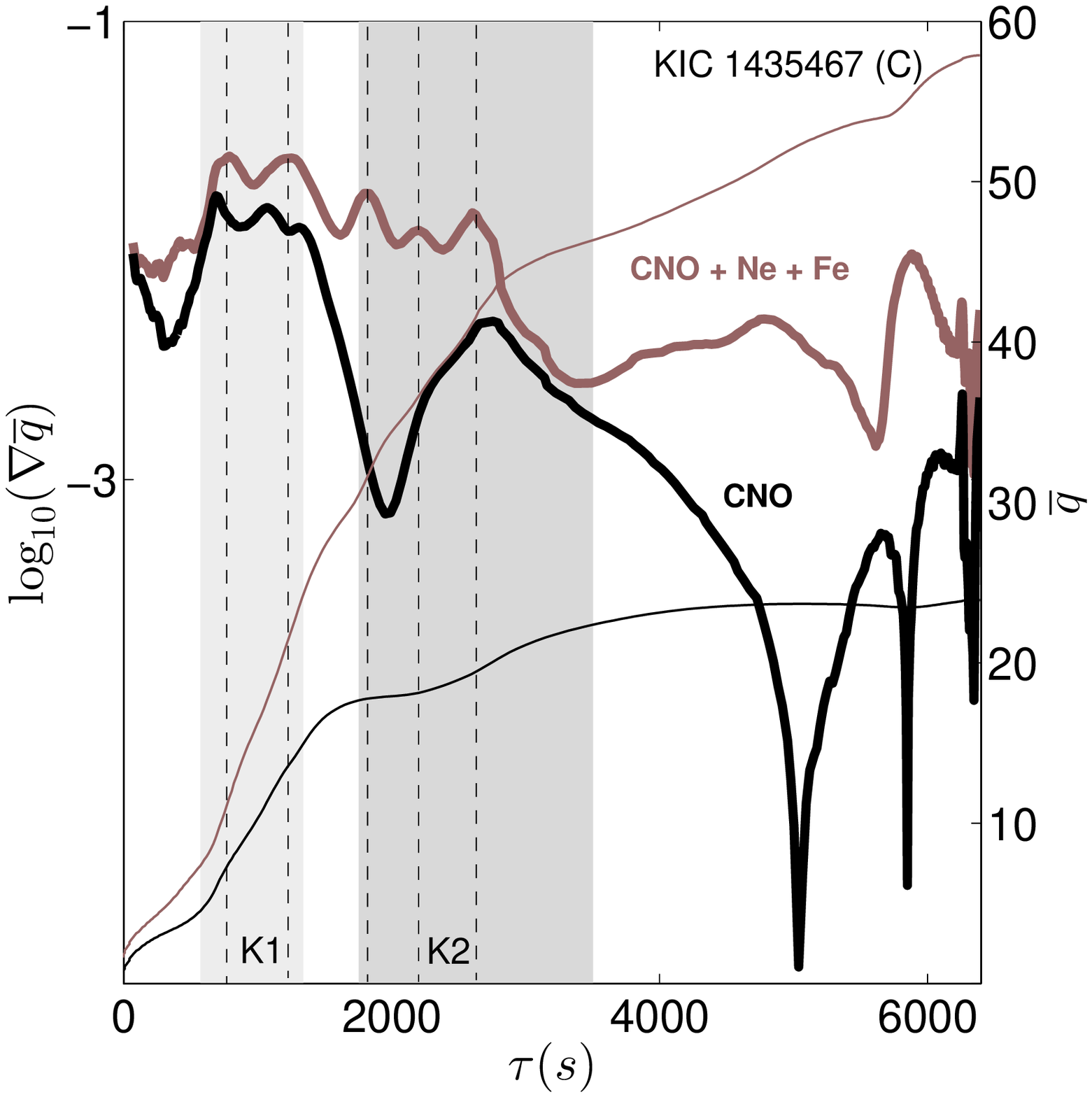}{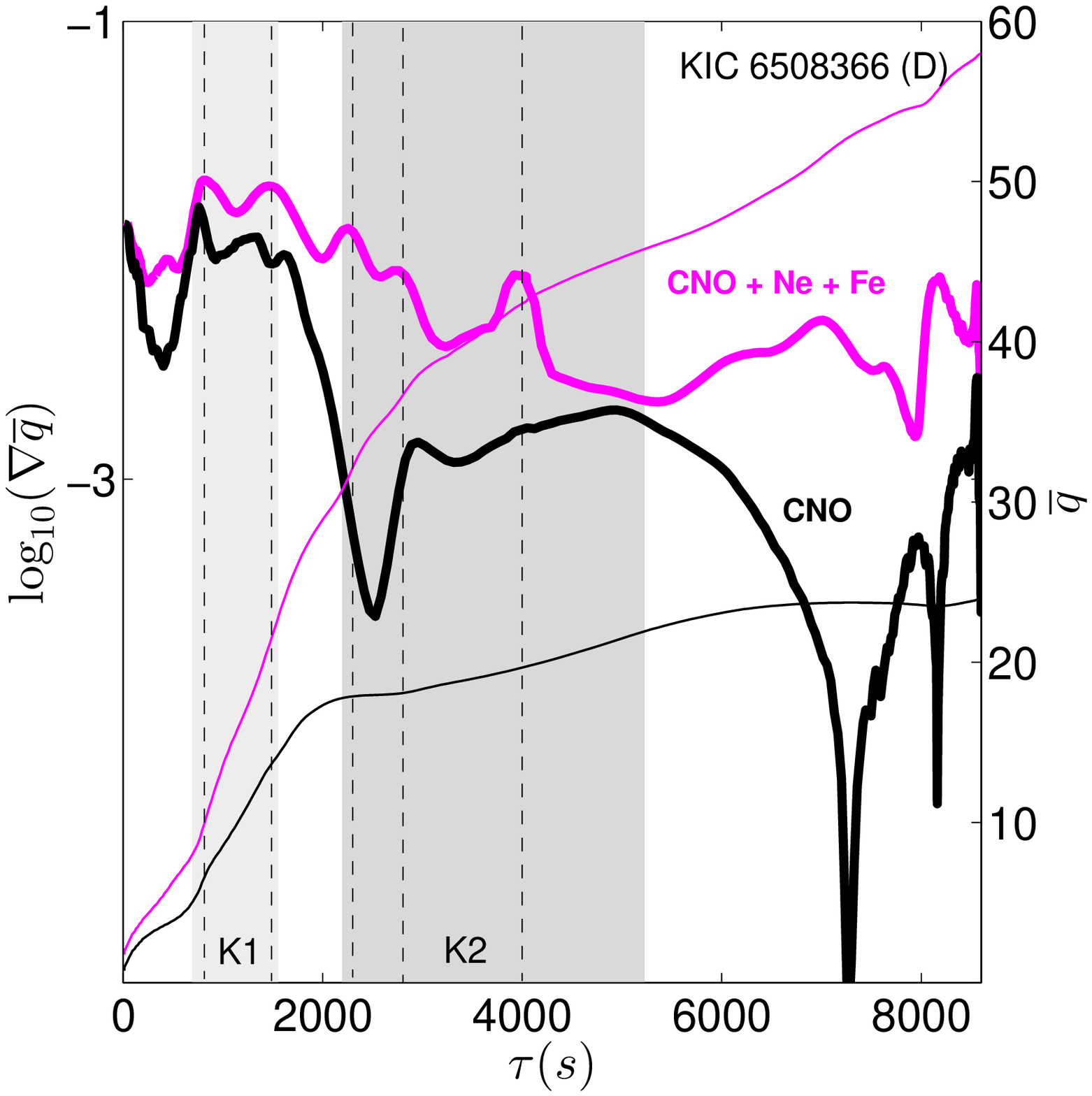}
	\caption{Effective mean ionic charge (thin solid lines) and the correspondent gradient (thick solid lines) plotted against the acoustic depth $\tau$ for each star. Colors represent the case when the mean effective ionic charge is defined by equation \ref{eq2}, i.e., taking into account seven chemical elements. Black solid lines do not include neon and iron. Vertical bars indicate the regions of abrupt variation of the effective mean ionic charge. Region K1 (light-gray bar) is mainly influenced by the ionization of light elements, whereas the variations of ionic charge in region K2 (medium-gray bar) are induced by the ionization of heavy elements. Dashed vertical lines indicate the positions of the peaks in the gradient of the charge.  We note that as the acoustic radius of a cool star (for instance, the left top panel) is significantly different from the acoustic radius of a hot star (for instance, the right bottom panel), the overall acoustic structure of all stars is consistent among them.}
\end{figure*}\label{fig2}

\begin{figure*}
	\plottwo{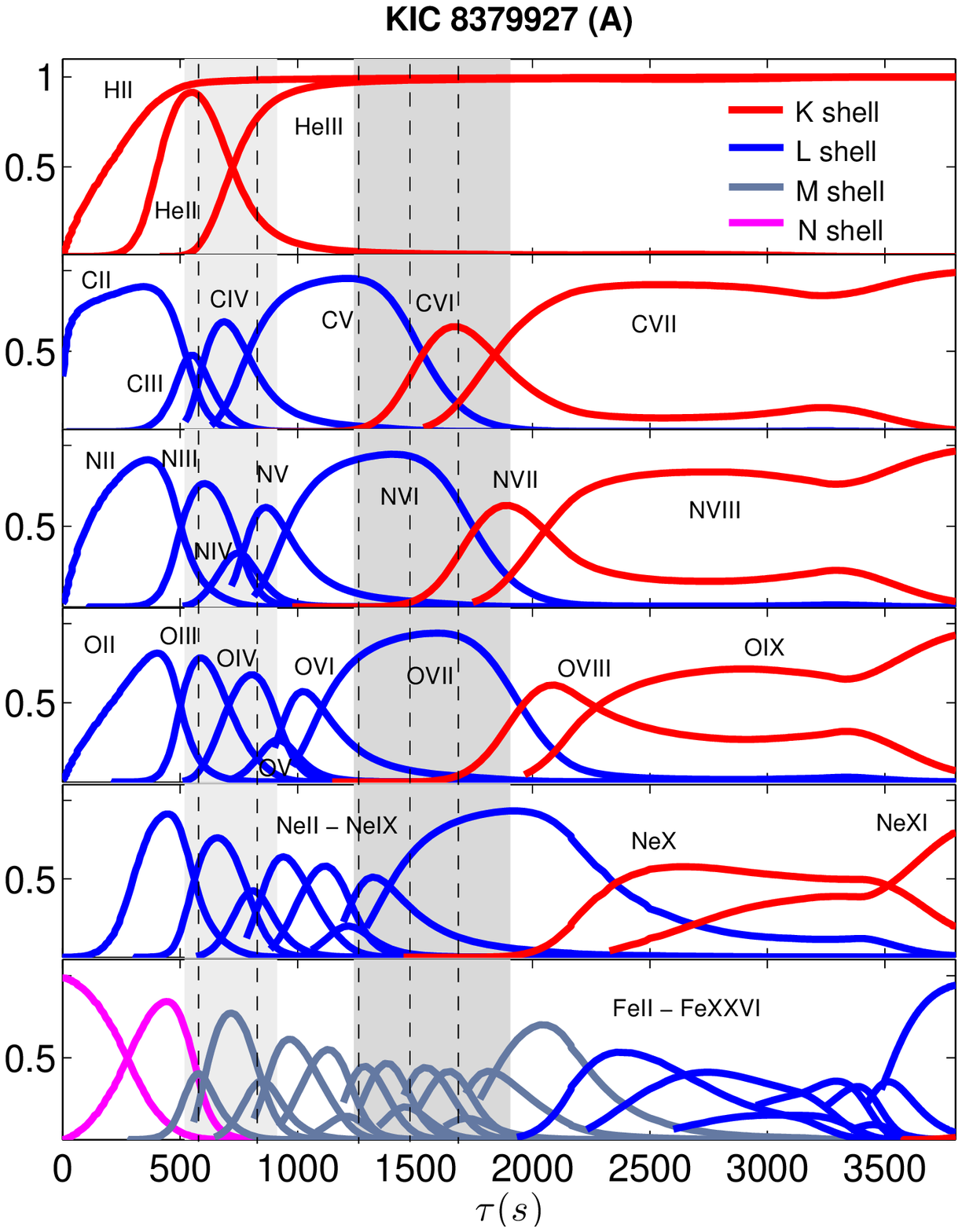}{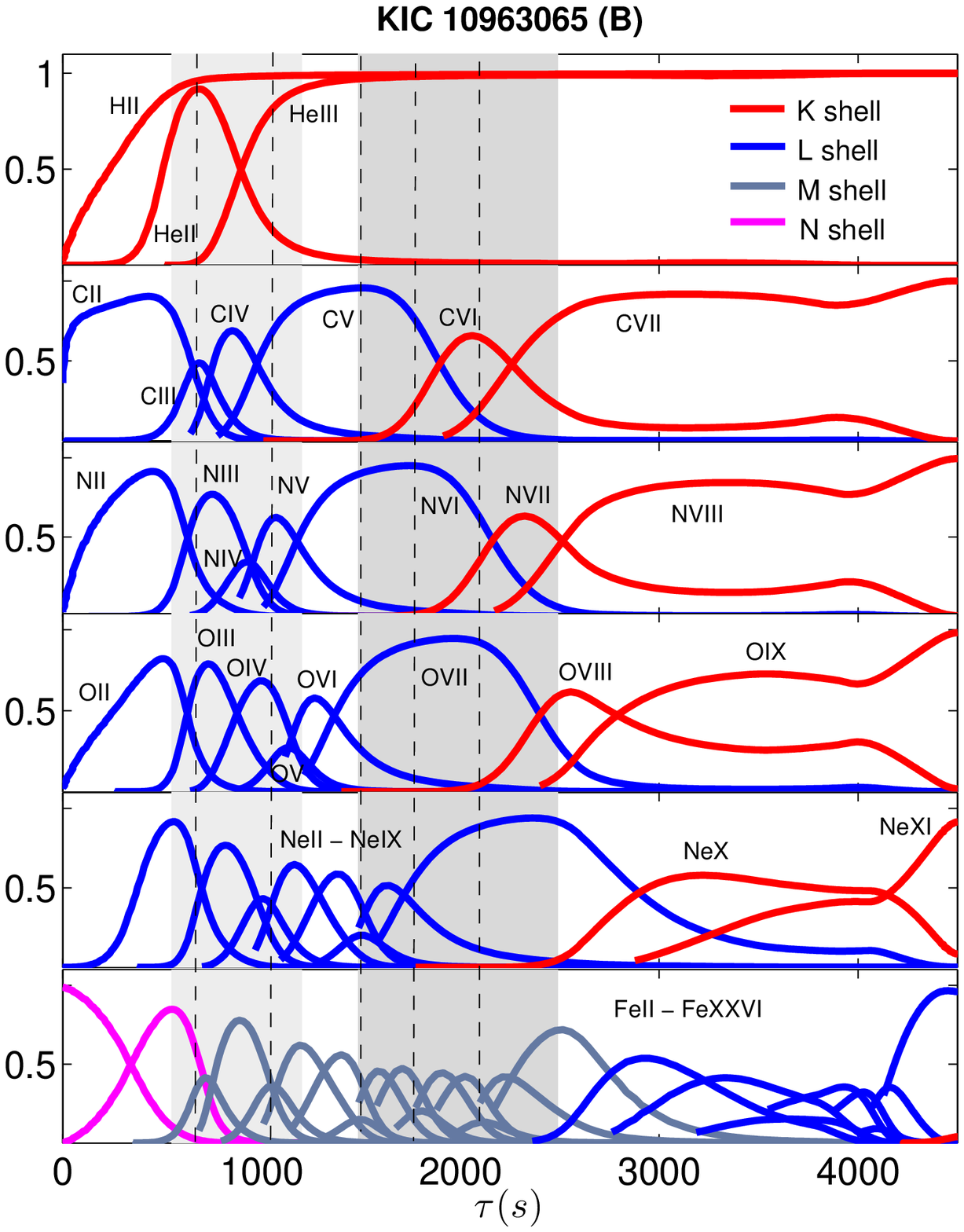}\\
	\plottwo{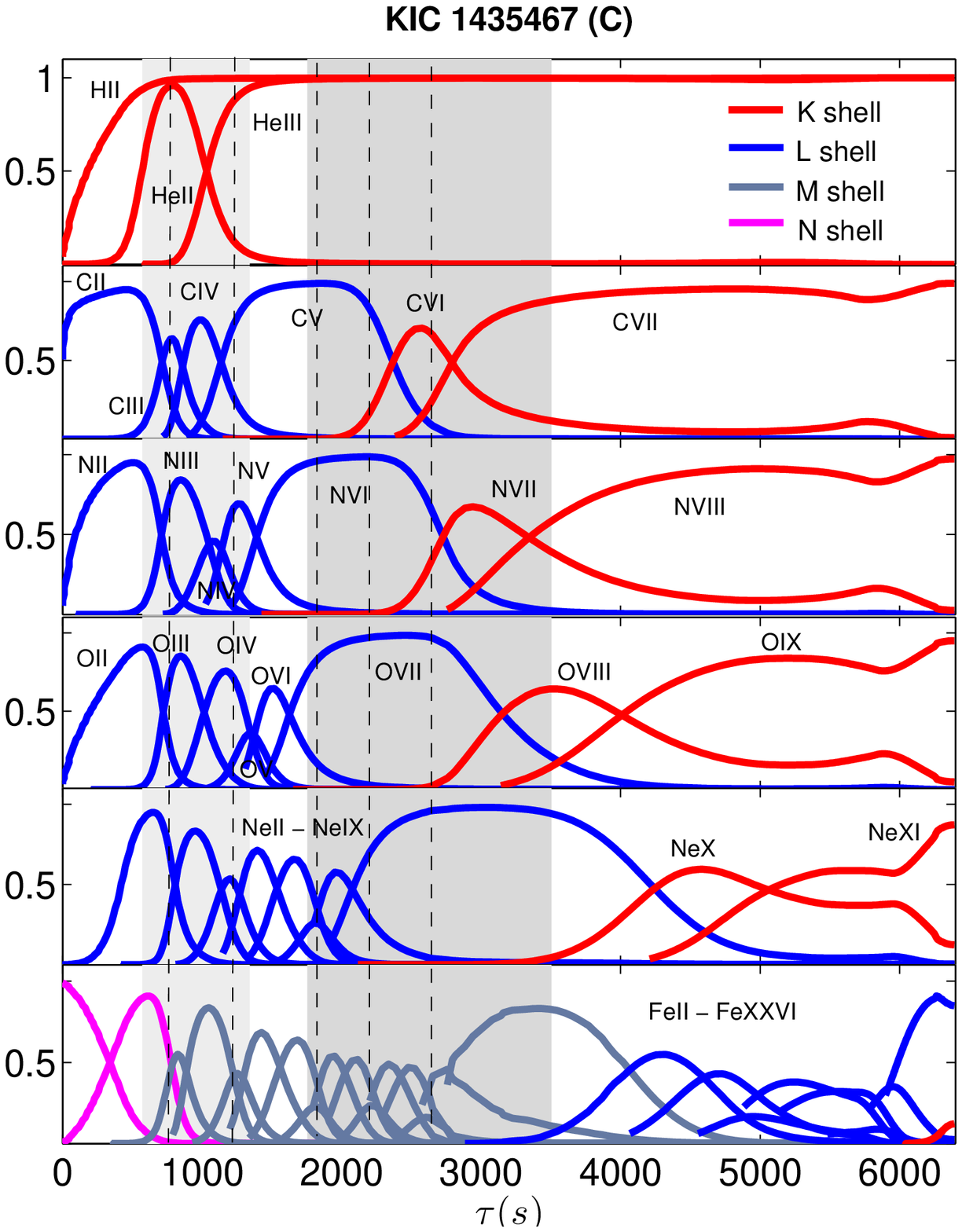}{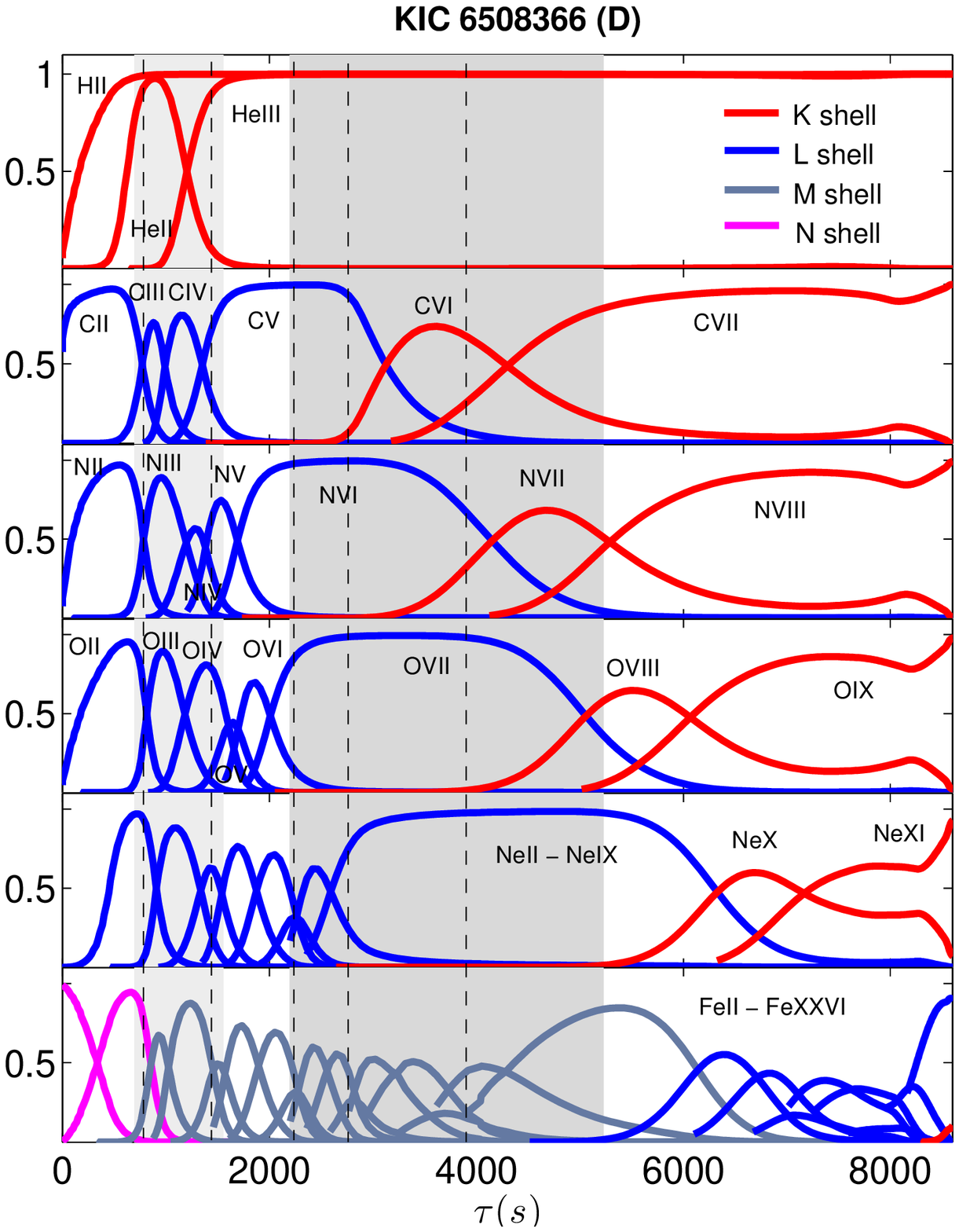}
	\caption{Degrees of ionization for hydrogen, helium, carbon, nitrogen, oxygen, neon, and iron plotted against the acoustic depth for all of the four stars. Vertical bars indicate the two regions of abrupt variation of the effective mean ionic charge. Region K1 is represented with a light-gray vertical bar and region K2 with a vertical medium-gray bar. Dashed vertical black lines indicate the locations of the relative maximums of the gradient of the effective mean ionic charge.}
\end{figure*}\label{fig3}

\section{The diversity of the Kepler sample} \label{sec2}

	\subsection{Asteroseismology of Solar-type Stars}

The observed oscillations in solar-type stars, which are essentially standing acoustic waves, have frequencies that are determined by the profile of the sound speed in the stellar interior. Asteroseismology provides the possibility to study the signatures of the regions of rapid variation of the sound speed. These regions are usually known as acoustic glitches, and have been being studied for a long time	\citep[e.g.][]{1988IAUS..123..151V, 1990LNP...367..283G, 1994A&A...282...73A,1999ASPC..173..257R, 2000MNRAS.316..165M, 2001A&A...368L...8M, 2004A&A...423.1051B}. The physical nature of the signatures is distinct. The signatures related to transitions between convective and radiative regions lead to discontinuities in the derivatives of the sound speed, whereas signatures of ionization zones do not usually lead to discontinuities in the sound speed derivatives, and instead produce a lowering of the first adiabatic exponent. The structural variation inside an ionization zone can be considered to be a sharp feature and produces a small shift in the eigenfrequencies, which in turn, is an oscillatory function of the frequency of the mode. Ionization zones can thus be studied using appropriate seismic methods, such as, the second differences $\delta_2 \nu_{\ell,n}=\nu_{\ell,n-1} - 2\nu_{\ell,n} + \nu_{\ell,n+1}$ \citep[e.g.,][]{2009arXiv0911.5044H, 2014ApJ...782...18M} or methods based on the derivative of the phase shift \citep[e.g.,][]{1991Natur.349...49V}.
The solar-type stars observed by {\it{Kepler}} with the highest S/N were recently reunited in the {\it{Kepler}} Legacy sample \citep{2017ApJ...835..172L, 2017ApJ...835..173S}. Figure \ref{fig1} shows an asteroseismic diagram with 20 stars of this sample. The stars in this figure cover a particularly large range of values of the large-frequency separations and surface gravities. Among these 20 stars, we highlighted four stars, which are our target stars in this work. They are representative of the diversity in the range of the observed values of the large-frequency separations, from $51.6$ to $120 \, \mu$Hz. Hence, we expect the four stars to be  reliable representatives of the \textit{Kepler} sample concerning to values of radii, effective temperatures and also luminosities.

\begin{figure*}
	\centering
	\includegraphics[width=0.53\textwidth]{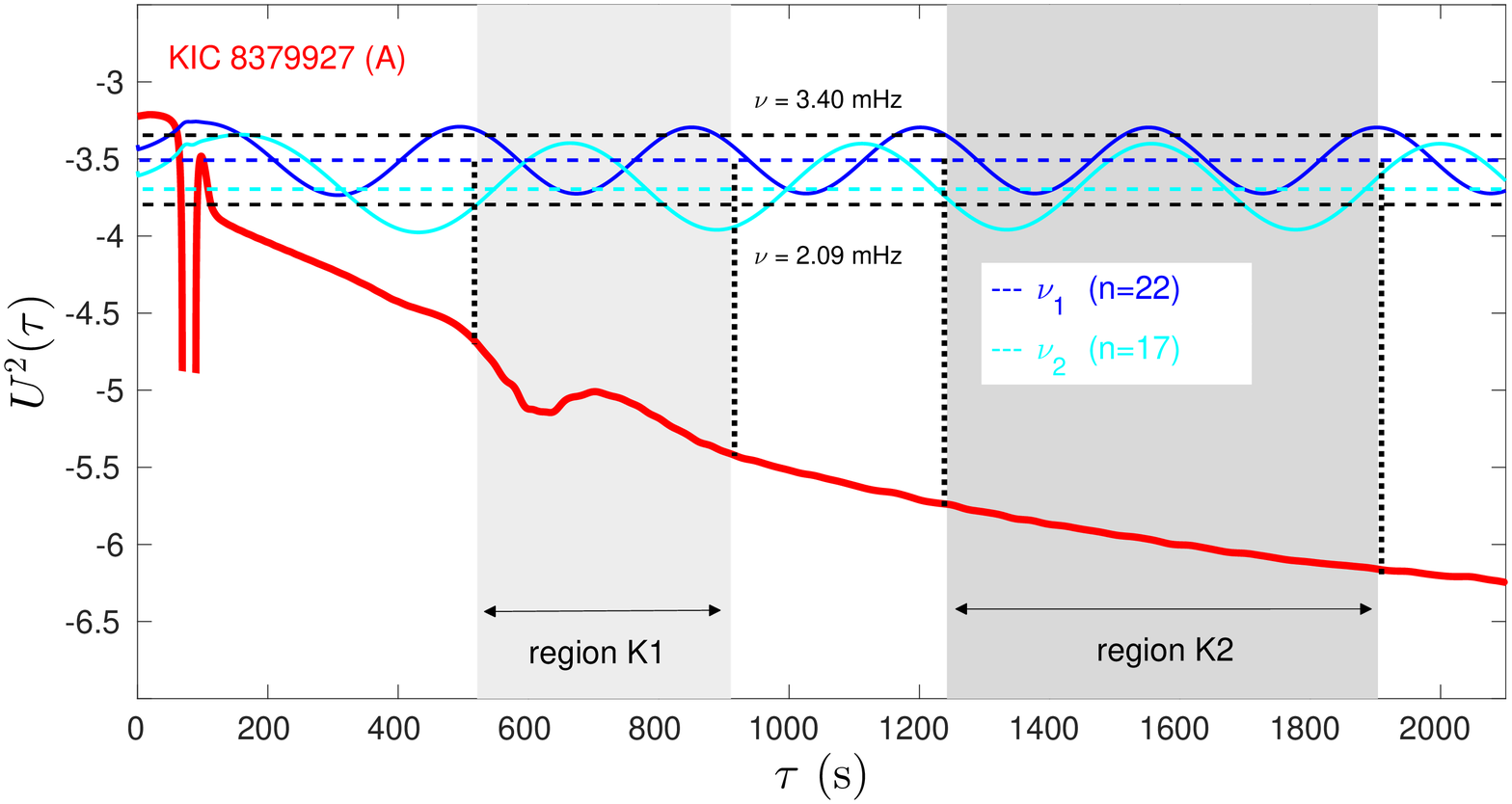}
	\includegraphics[width=0.265\textwidth]{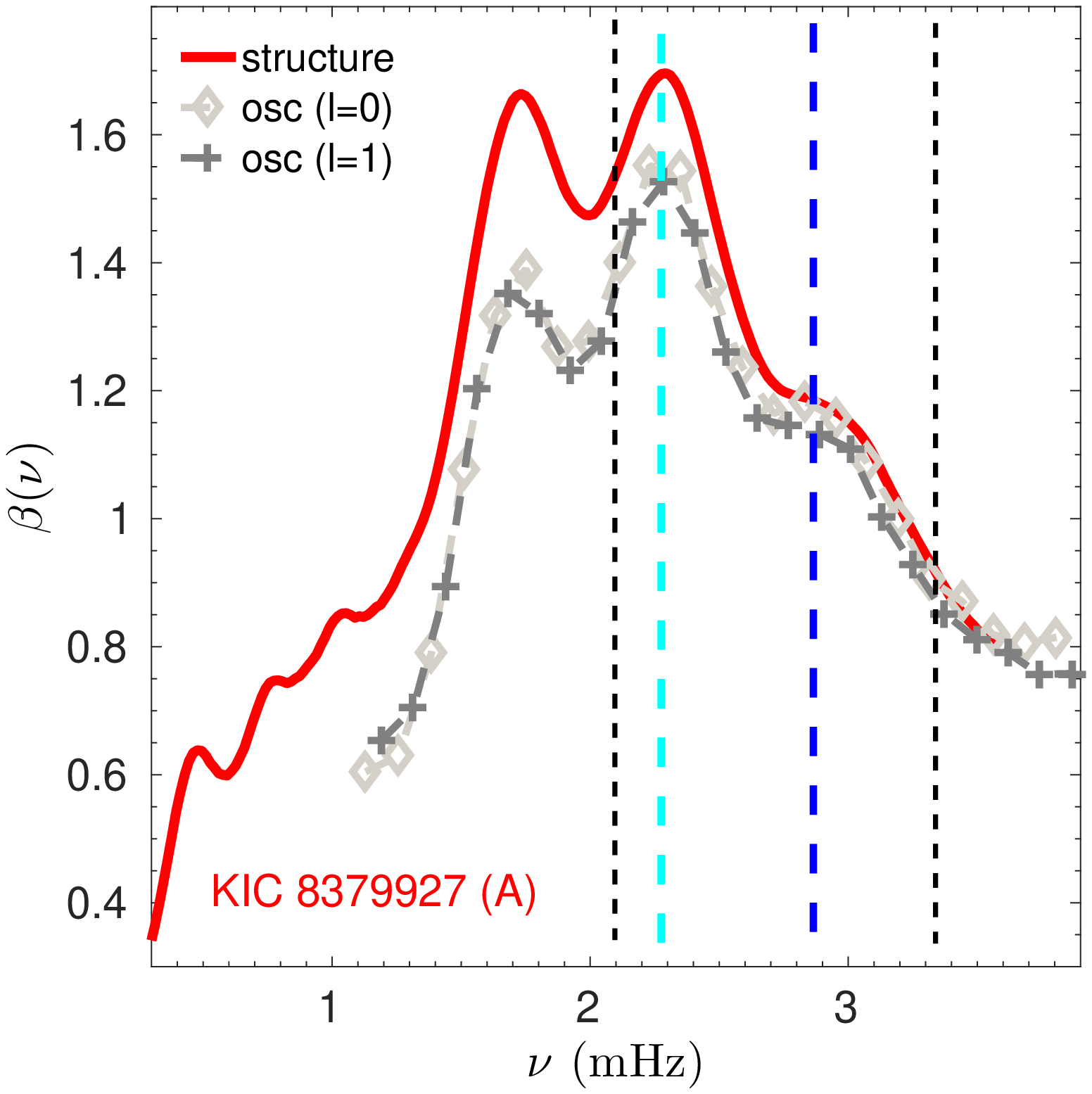}\\
	\includegraphics[width=0.53\textwidth]{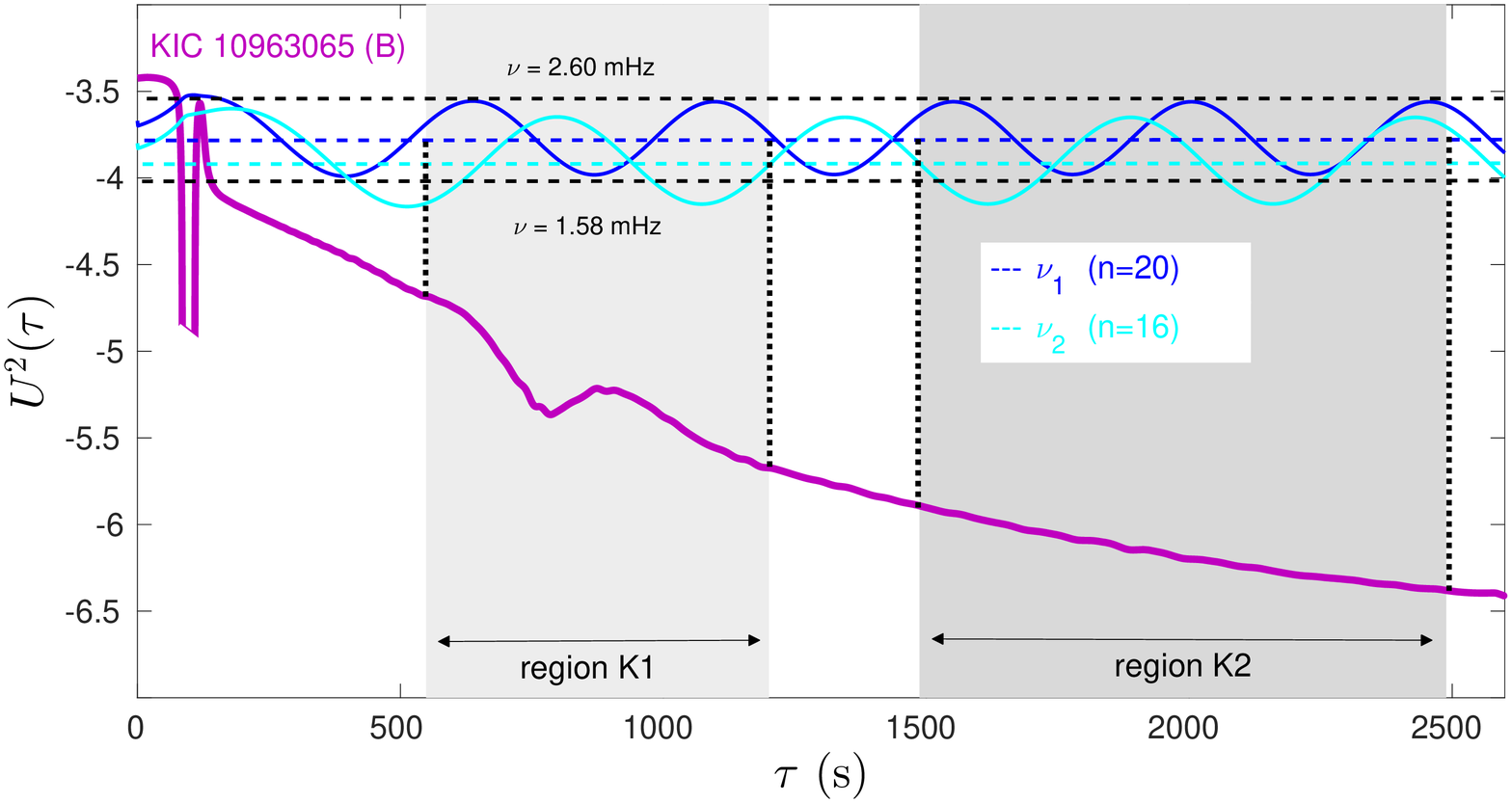}
	\includegraphics[width=0.265\textwidth]{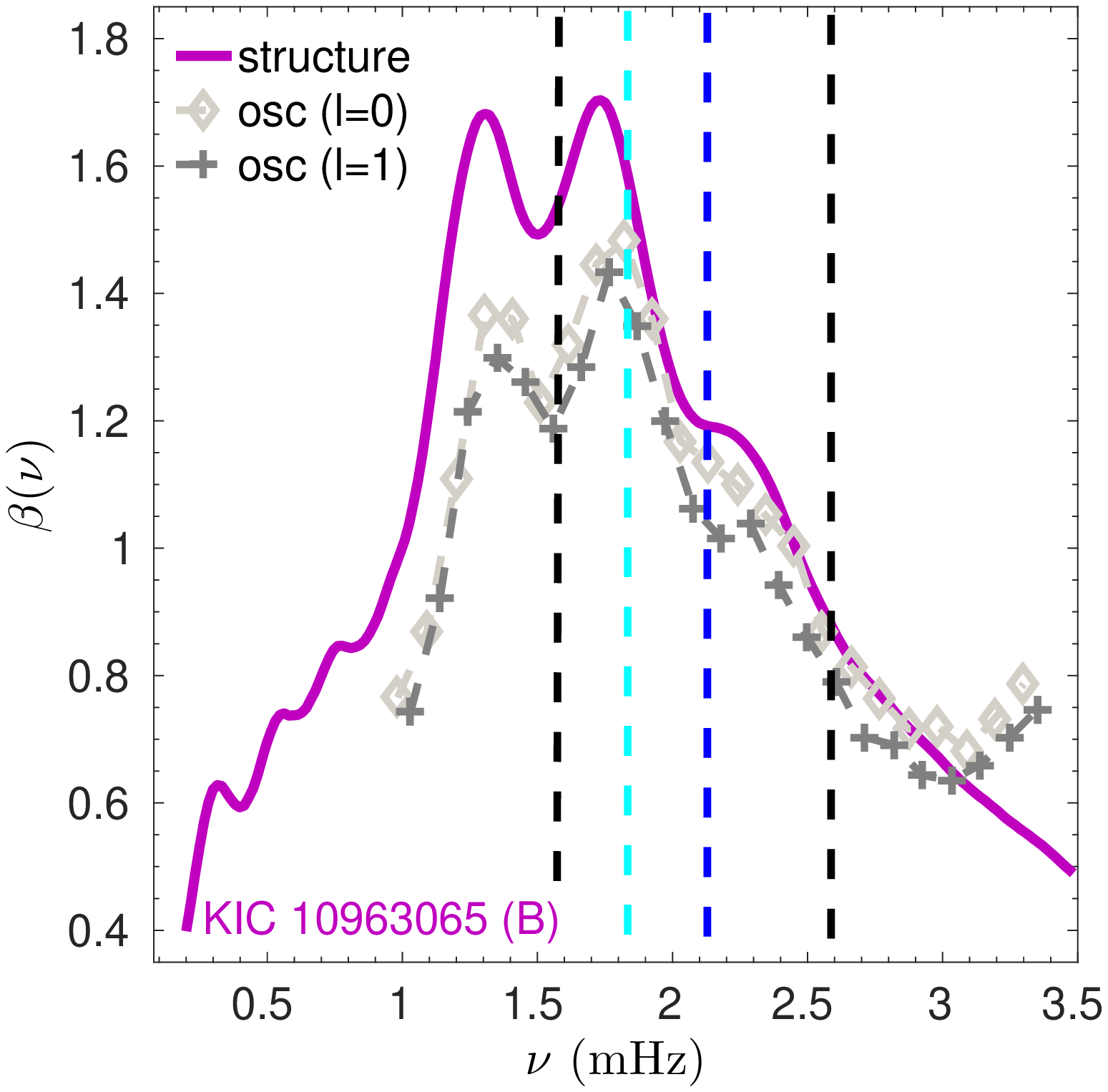}\\
	\includegraphics[width=0.53\textwidth]{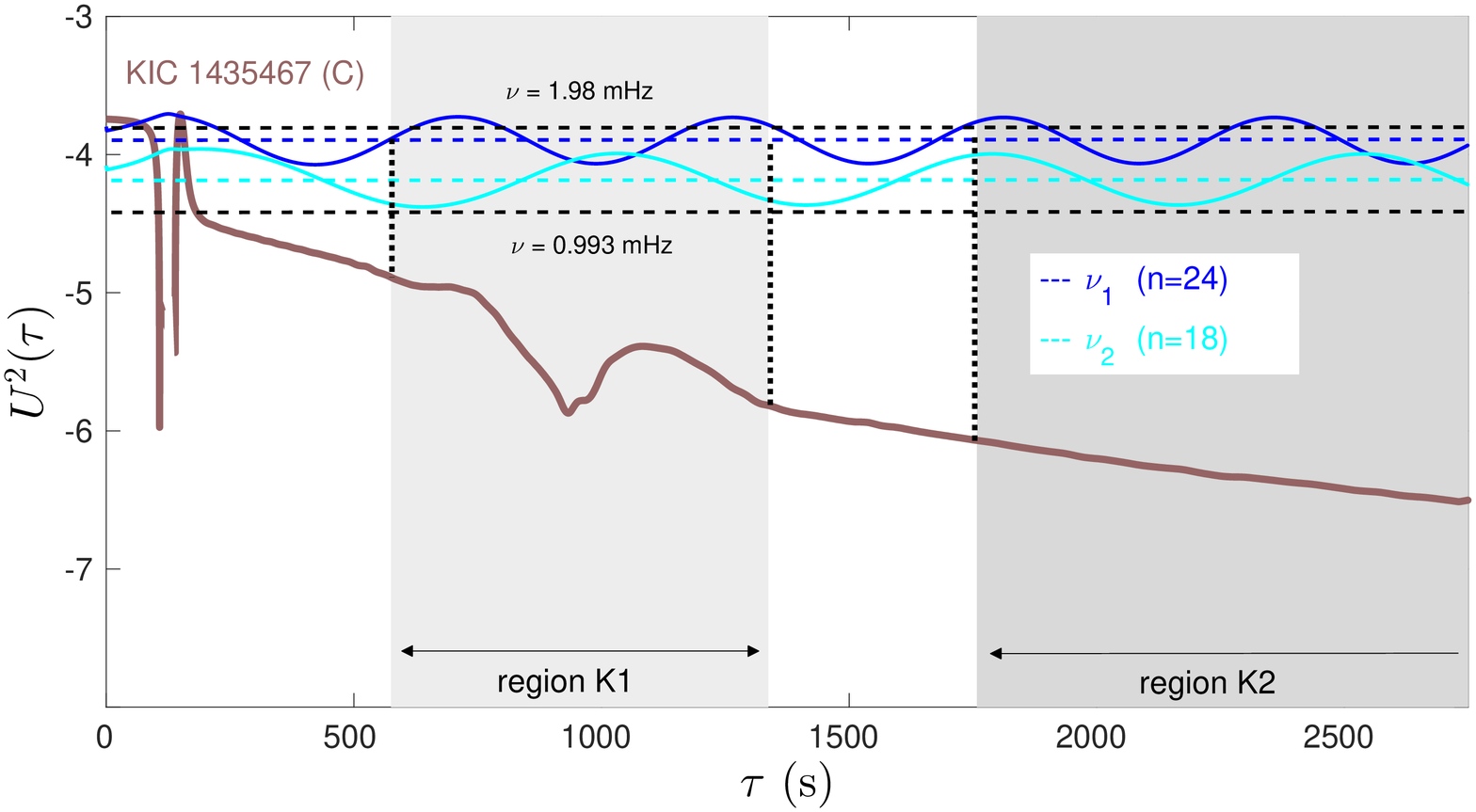}
	\includegraphics[width=0.265\textwidth]{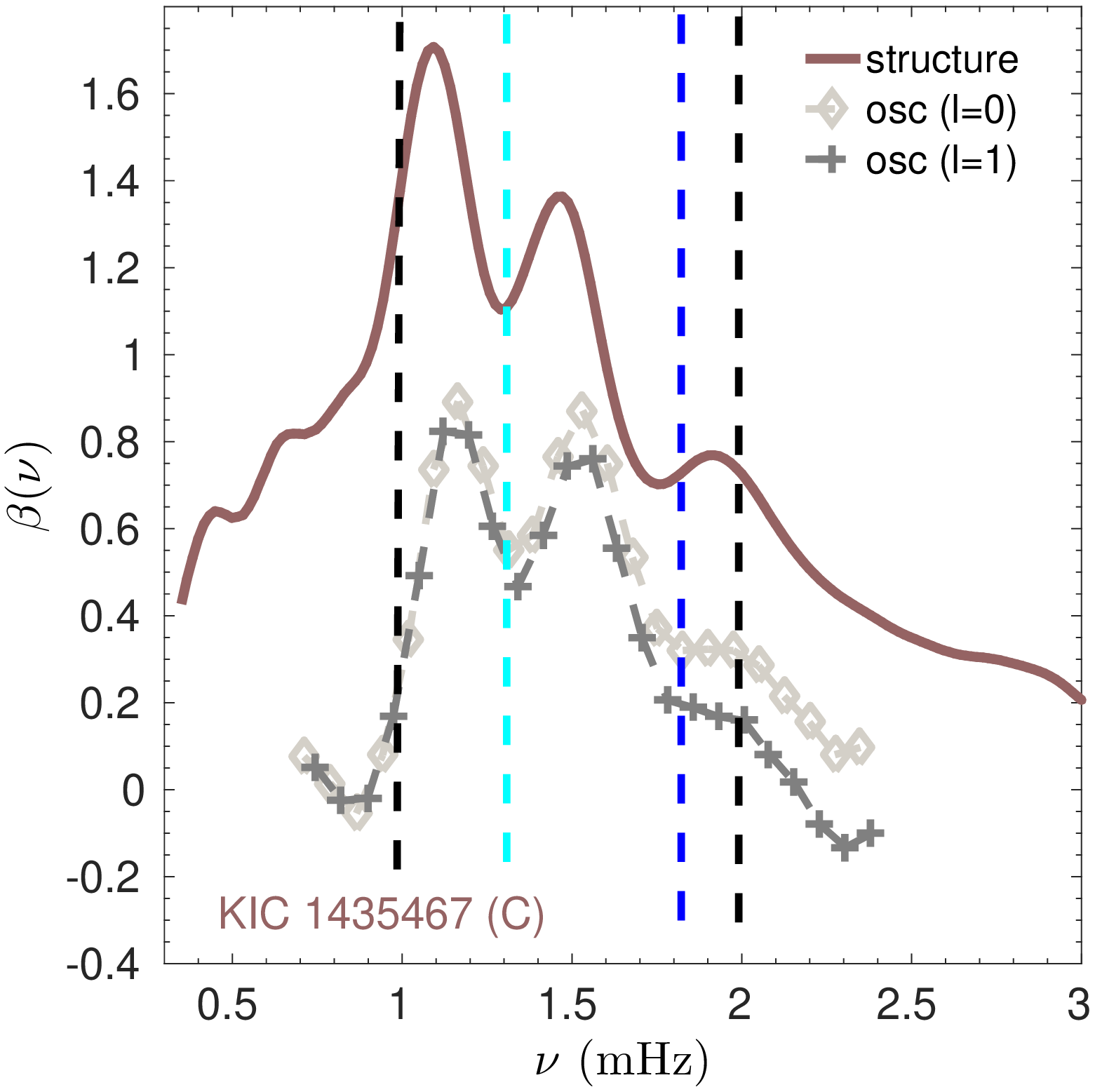}\\
	\includegraphics[width=0.53\textwidth]{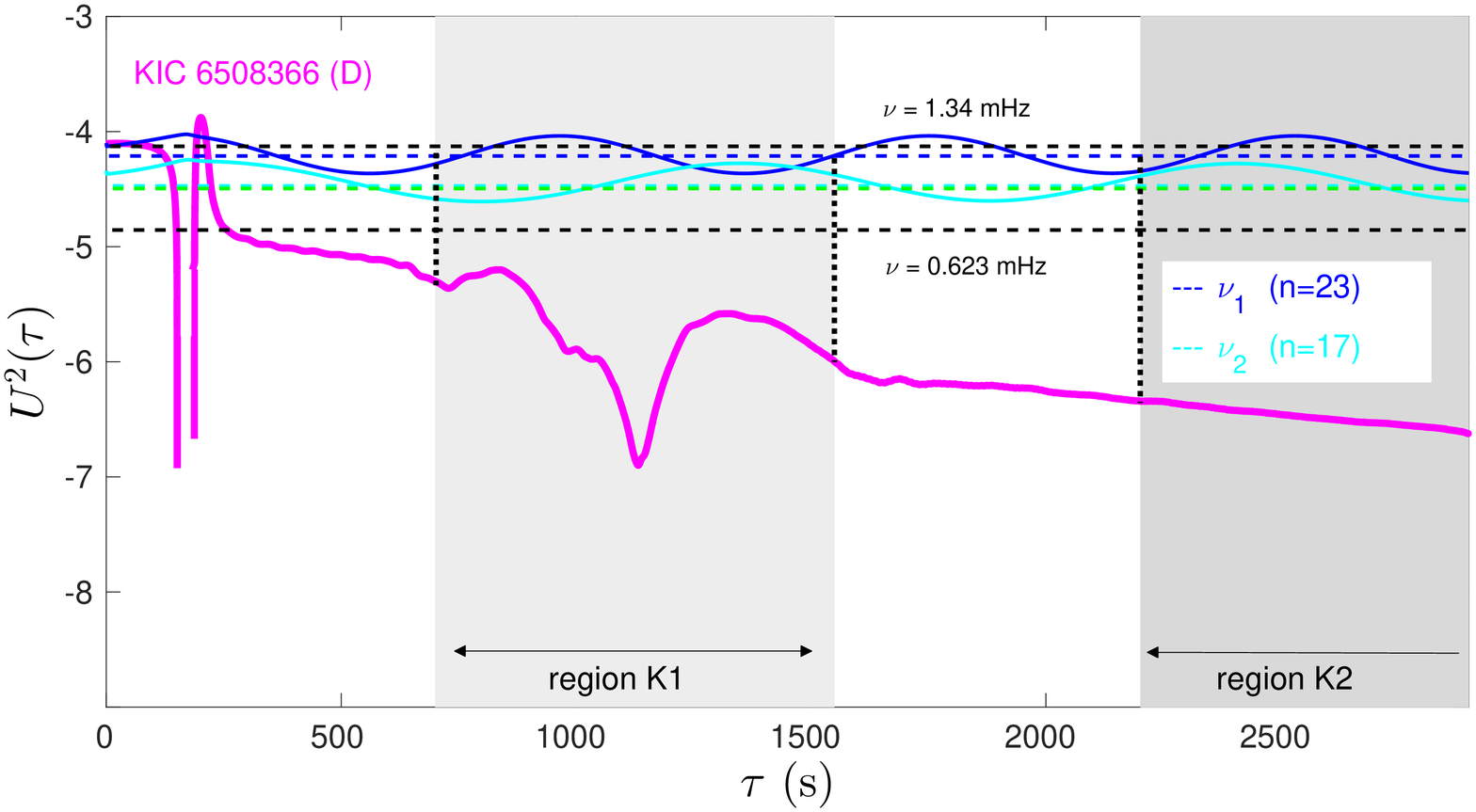}
	\includegraphics[width=0.265\textwidth]{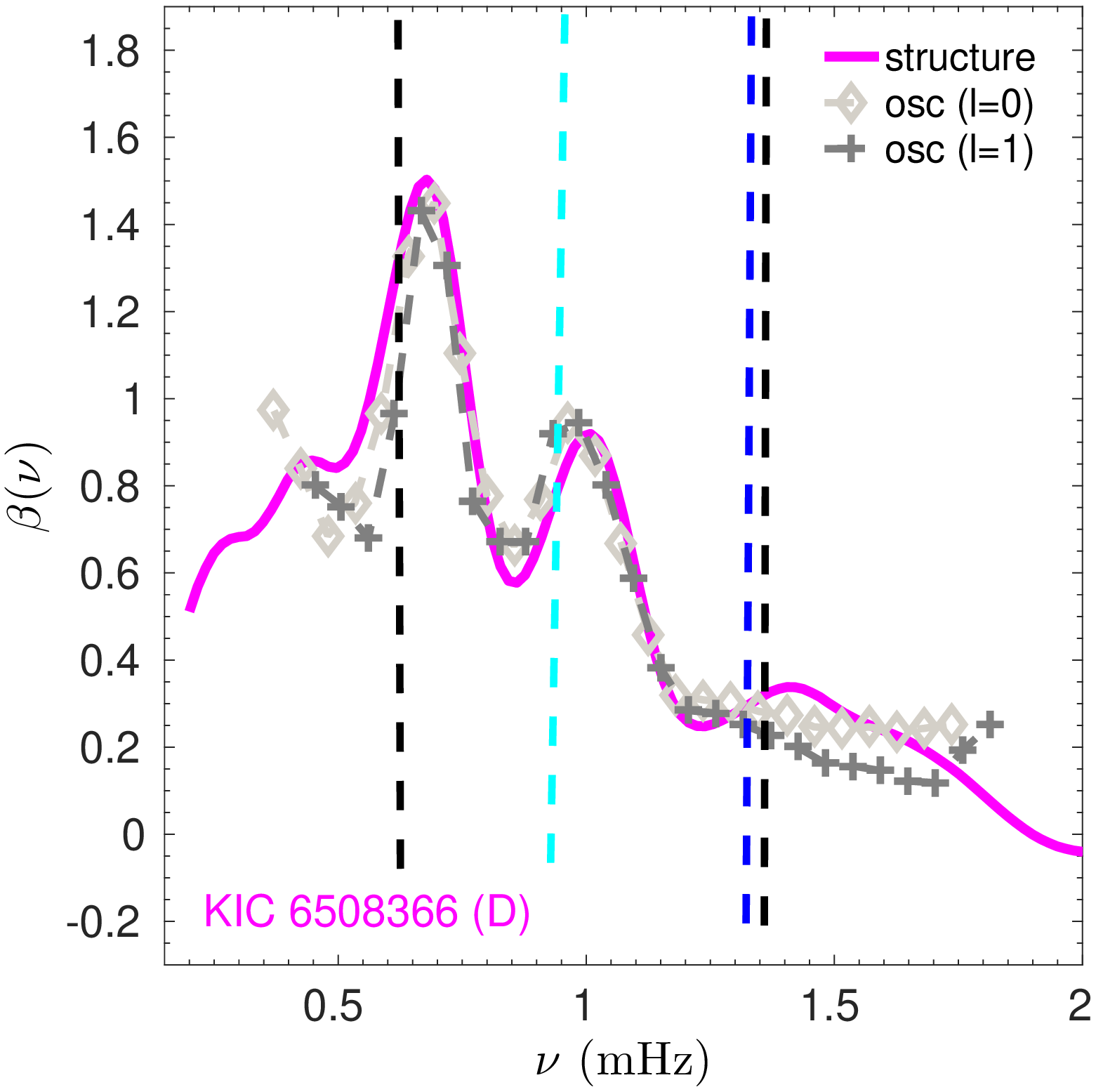}
	\caption{Left: reflecting acoustic potentials for the four selected stars. Horizontal dashed lines represent the squared values of angular frequencies, $\omega^2=(2\pi \nu)^2$: black color represents the limits of the observational window, and blue and cyan represent two reference frequencies defined within the observational window. For these two frequencies, $\nu_1$ and $\nu_2$, we also represented the wave functions with the corresponding amplitudes scaled arbitrarily. Right: the solid line is the seismic parameter $\beta(\nu)$ computed from the structure of the theoretical model envelope. Gray lines and symbols: $\beta(\nu)$ computed from a table of theoretical frequencies for the degrees $l=0$ and $l=1$. Vertical dashed lines indicate the position of the frequencies represented in the acoustic potentials. Despite the different acoustic depths for the locations of the base of the convective zones, the overall theoretical acoustic structure of all stars is not strikingly different.}
\end{figure*}\label{fig4}

\subsection{The modeling of the benchmark stars}

Stars were modeled with the stellar structure and evolution code CESAM \citep{1997A&AS..124..597M}. The theoretical models computed use the following input physics: OPAL  as equation of state \citep{2002ApJ...576.1064R} and the most recent OPAL opacities; nuclear reaction rates from the NACRE compilation \citep{1999NuPhA.656....3A}; convection is described by the mixing-length formalism \citep{1958ZA.....46..108B}; diffusion was included according to Burgers theory \citep{1969fecg.book.....B}; and the solar mixture of \citet{2009ARA&A..47..481A}. The atmosphere was characterized according to Eddington's gray law and the contact with the upper part of the envelope occurs at an optical depth of 20. 
For each star, we compute a broad grid of stellar models. These grids are obtained by varying three parameters: the stellar mass, $M$, the initial metallicity, $Z$, and the mixing-length parameter, $\alpha$. The breadth of the grids spans values with physical significance in each case. From all of the generated stellar models, we choose those that reproduce the values of the observational constraints in Table \ref{table:1} within the error bars (1$\sigma$ for the non-seismic constraints and 2$\sigma$ for the large separation). For all of the obtained valid models, the theoretical acoustic oscillation frequencies were computed with the ADIPLS code \citep{2008Ap&SS.316..113C}. Finally, the age is fine-tuned through a $\chi^2$ minimization to reproduce all observational constraints in Table \ref{table:1} as closely as possible. The method is similar to that used by \citet{2013ApJ...765L..21C} and \citet{2017ApJ...843...75B}.

The resulting model parameters are given in Table \ref{table:2}. We found that our models for these benchmark stars are identical to the models found in the literature \citep[e.g.,][]{2017A&A...601A..67C}.

\section{Ionization in the Deep Convective Zone of Solar-type Stars}  \label{sec3}

	\subsection{The effective mean ionic charge and the corresponding gradient}

As the temperature increases with depth, toward the center of the star, atoms begin gradually to lose electrons. When temperature is high enough they can reach full ionization. Light elements like hydrogen and helium are the first to lose all the electrons and to reach a state of full ionization. In solar-type stars this is well known to happen approximately around 2\% under the surface of the stars. Nevertheless, partial ionization of heavier elements will occur throughout larger depths.

Considering that the collisional processes dominate in the interiors of solar-type stars, the Saha equation \citep{1921RSPSA..99..135S} can describe the ratio between different ionization states for a given position inside the star. Specifically, it can be used to compute the mean ionic charges of each element at different depths inside the star. Following \citet{2017MNRAS.466.2123B} we define a mean effective ionic charge
\begin{equation} \label{eq1}
	\overline{q}=\sum_i \langle Q_i \rangle
\end{equation}
where $\langle Q_i \rangle$ is the mean ionic charge of each chemical element considered. The idea is to describe the ionization processes occurring in the theoretical envelopes of the stellar models of the four benchmark F-stars from a viewpoint of the variation of an effective ionic charge.  We intend to measure the efficacy of ionization at a certain depth inside the star by computing the variations of the mean ionic charge for a given set of chemical elements. The set of elements considered in this work, in addition to the light elements hydrogen and helium, includes five heavy elements: carbon, nitrogen, oxygen, neon and iron. The mean effective ionic charge is then defined as
\begin{eqnarray} \label{eq2} \nonumber
	\overline{q}= \langle Q_{\text{H}} \rangle + \langle Q_{\text{He}} \rangle +\\ + \langle Q_{\text{C}} \rangle + \langle Q_{\text{N}} \rangle + \langle Q_{\text{O}} \rangle +  \langle Q_{\text{Ne}} \rangle + \langle Q_{\text{Fe}} \rangle .
\end{eqnarray}
The relevance of a partial ionization region can be characterized by the variation rate of the mean effective ionic charge in this region. To quantify this variation, we take the gradient of the effective mean ionic charge,
\begin{equation} \label{eq3}
\nabla(\overline{q}) \equiv \frac{d\,\overline{q}}{d\,\tau},
\end{equation}
with respect to the acoustic depth,
\begin{equation} \label{eq4}
\tau = \int_{r}^{R}\frac{dr}{c} \, .
\end{equation}
Here, $R$ is the radius of the star, and $c$ is the adiabatic sound speed. Figure \ref{fig2} shows the effective mean ionic charges and the respective gradients for all the theoretical models. Confining our analysis to the outer layers of the stars, we can identify two regions of abrupt variation of the ionic charge. The first region is identified as region K1 in Figure \ref{fig2} and corresponds to a region of rapid rise of $\overline{q}$. It is represented by a light-gray vertical bar and correlates to the highest values of the gradient $\nabla (\overline{q})$. The location of this region K1 coincides, in all stellar models, with the region usually known as the region of the second ionization of helium. Located deeper, but still inside the convective zones of the stellar models, we highlighted with a medium-gray vertical bar a second region: region K2 in Figure \ref{fig2}, which also corresponds to a region of rapid increase (but not so rapidly as in region K1) of $\overline{q}$.  This region correlates to a series of second-highest peaks of the gradient of the mean effective ionic charge. Because light elements, hydrogen and helium, are already completely ionized at these depths, these peaks in the gradient are originated by ionization processes related to heavy chemical elements.

	\subsection{Ionization Fractions: The Regions of the K-Shell Ionization}
	
To better understand the two highlighted regions in Figure \ref{fig2}, and also to describe the successive ionizations occurring from the surface toward the interior of the star, we have computed the ionization fractions for the seven chemical elements: hydrogen, helium, carbon, nitrogen, oxygen, neon, and iron. Figure \ref{fig3} shows these for all of the models.	
The relevant ionization region closer to the surface is indeed related to the second ionization of helium in all stellar models. However, this is a broad region with contributions from another ionization processes, such as the first ionization of helium and the loss of valence electrons in heavier elements. The second identified ionization region is exclusively induced by the ionization of carbon, nitrogen, oxygen, neon, and iron. This is because  at this depth, hydrogen and helium are completely ionized and cannot contribute to the variation of the ionic charges. Moreover, the distribution of the ionization fractions in Figure \ref{fig3} allows us to associate this ionization region with the loss of the electrons of the K-shell for the elements of the second period of the periodic table. K-shell ionizations of heavier elements like iron occur much deeper inside the star.  The other ionization region, located closer to the surface, can also be seen as a K-shell ionization region; this is the region where helium loses its only two electrons (K-shell electrons).
The two locations, in the convective zones of these four models, where the variation of the ionic charge is the most abrupt, can be both related to K-shell ionizations.
K1 is the closest to the surface and related to the ionization of light elements, whereas K2 is located deeper and related only with the ionization of heavy elements.

Finally, we note that in Figure \ref{fig3} the ionization states \ion{C}{5}, \ion{N}{6}, \ion{O}{7}, \ion{Ne}{9}, and \ion{Fe}{17} prevail over a large plateau. All the five are ions with a noble gas configuration, hence with high values for the ionization potentials. Ions with such noble gas configurations are more stable in relation to the others, and stay in the same degree of ionization for a large range of temperatures. They also play an important role in the variations of the ionic charges.

%%%%%%%%%%%%%%%%%%%%%%%%%%%%%%%%%%%%%%%%%%%%%%%%%%%%%%%%%%%%%%%%%%%%%%%%%%%%%%%%%%%%%%%%
%%%%%  TABLE 3
%%%%%%%%%%%%%%%%%%%%%%%%%%%%%%%%%%%%%%%%%%%%%%%%%%%%%%%%%%%%%%%%%%%%%%%%%%%%%%%%%%%%%%%%
\begin{deluxetable*}{ccccccc} 
	\tablecaption{Comparison of surface and initial abundances for all the models \label{table:3}}
	\tablecolumns{7}
	\tablenum{3}
	\tablewidth{0pt}
	\tablehead{
		\colhead{Star Id.} &
		\colhead{$Z_s$} &
		\colhead{$Z_0$} &
		\colhead{$Z$ Reduction (\%)} &
		\colhead{$Y_s$} &
		\colhead{$Y_0$} &
		\colhead{$Y$ Reduction (\%)} 
	}
	\startdata 
	A & 0.01799 & 0.0196 &8.20& 0.2377 & 0.2640 &9.90  \\
	B & 0.01287 & 0.0150 &16.5& 0.2153 & 0.2600 &17.2  \\
	C & 0.01967 & 0.0254 &22.5& 0.1910 & 0.2650 &27.9  \\
	D & 0.01368 & 0.0201 &31.9& 0.1716 & 0.2776 &38.1  \\
	\enddata
	%\tablenotetext{}{(a) \citet{2017ApJ...837...47V}}
\end{deluxetable*}
%%%%%%%%%%%%%%%%%%%%%%%%%%%%%%%%%%%%%%%%%%%%%%%%%%%%%%%%%%%%%%%%%%%%%%%%%%%%%%%%%%%%%%%%
%%%%%  TABLE 3 - END
%%%%%%%%%%%%%%%%%%%%%%%%%%%%%%%%%%%%%%%%%%%%%%%%%%%%%%%%%%%%%%%%%%%%%%%%%%%%%%%%%%%%%%%%

\section{The Ionization Processes and the Oscillation Spectrum} \label{sec4}

The processes associated with the microphysics in stellar interiors have specific characteristics that are interconnected with the thermodynamics of the plasma. Among all the thermodynamics variables, the first adiabatic exponent, $\Gamma_1$, emerges as one of the most fundamental physical parameters  for understanding the behavior of acoustic oscillations in solar-type stars. The structure of the lower part of the convection zone depends on $\Gamma_1$, which follows an adiabatic stratification. Therefore, the structure of a major part of the convection zone is determined by the degree of ionization of the stellar matter. Partial ionization produces a lowering of $\Gamma_1$ values. Specifically, the depression induced in the first adiabatic exponent by the second ionization of helium has been extensively studied and discussed in the solar case \citep[e.g.,][]{1994MNRAS.268..880R,1995MNRAS.276..283C, 2005MNRAS.361.1187M}. Recently, it was also discussed in the case of solar-type stars \citep{2014ApJ...794..114V}.
The zone of the second ionization of helium has always deserved a special attention regarding the study of the acoustic glitches \citep[e.g.,][]{1990SvAL...16..108B, 1990LNP...367..283G,1994A&A...282...73A, 2004MNRAS.350..277B, 2007MNRAS.375..861H, 2011MNRAS.418L.119H}. The main reason of the focus on the \ion{He}{3} signature is the difficulty in isolating the hydrogen and the first ionization of helium zones, since they overlap. Using different methods, the signature of this second ionization of helium has been being used to infer about the helium abundance in the solar \citep{1991Natur.349...49V, 1995MNRAS.276.1402B} and solar-type envelopes \citep{2004MNRAS.350..277B, 2014ApJ...790..138V}. Effects due to ionization of heavier elements in the frequencies were studied by e.g., \citet{1998ESASP.418..543T}, \citet{2006ApJ...644.1292A}, and \citet{2008PhR...457..217B}. These effects are found to be generally fainter than effects due to helium ionization. More recently, \citet{2017MNRAS.466.2123B} unveiled a possible signature of heavy elements ionization in a solar-type star.

\begin{figure*}
	\plottwo{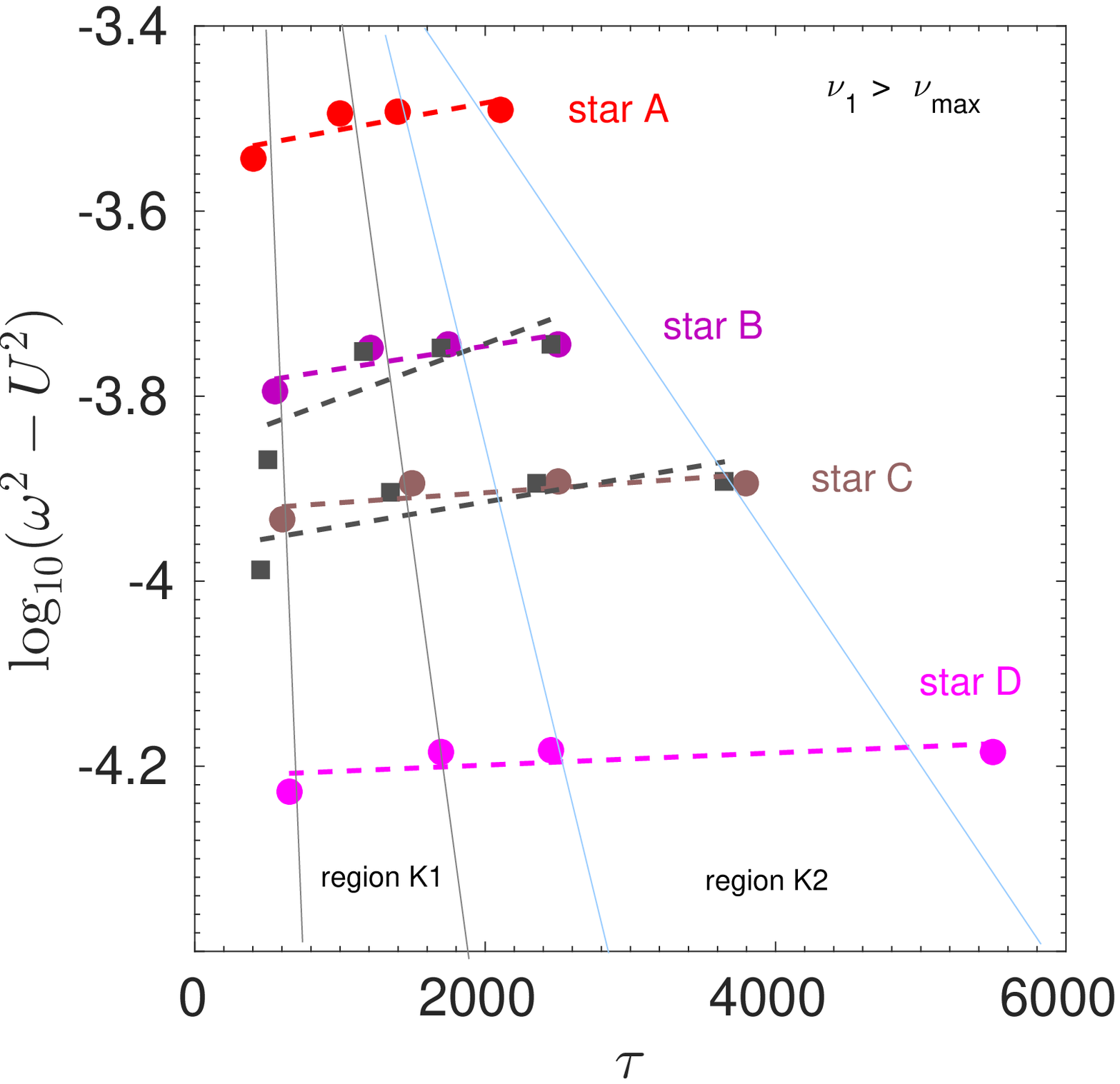}{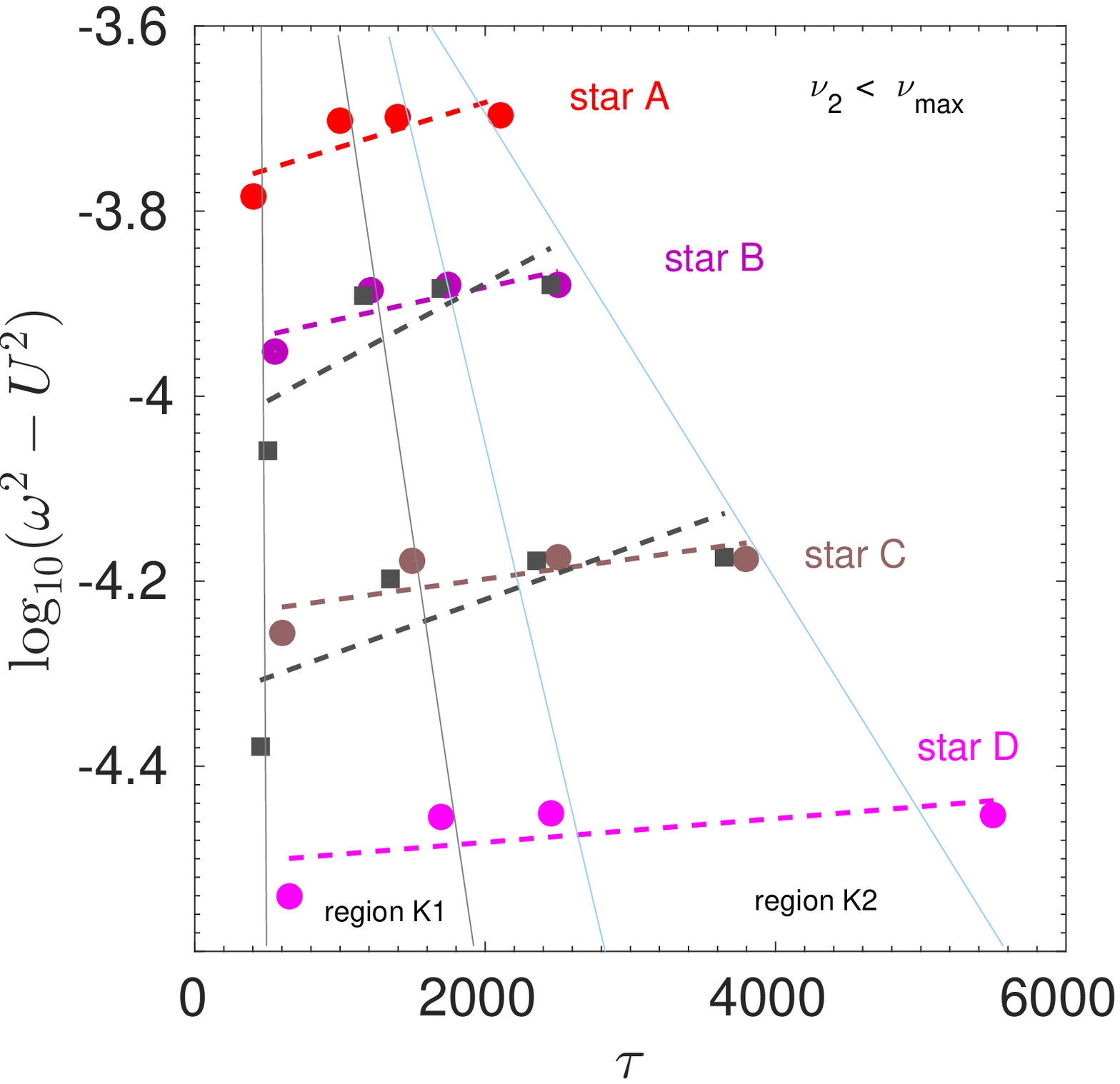}
	\caption{Left: the differences $\omega^2 - U^2$ represented for the frequency $\nu_1$ ($\nu_1$ is represented in Figure \ref{fig4}).
	These distances were measured for four different locations for each star. The locations are the limiting acoustic depths of regions K1 and K2. Each star is identified by the respective color. "Quasi-horizontal" dashed lines represent a linear fit to the distances $\omega^2 - U^2$ for each star. A large value of the slope of these lines indicates that the two regions (K1 and K2) influence the acoustic mode distinctively. The larger values of the slopes are obtained in stars A and B (cooler stars) meaning that in hotter stars, the influence of region 2 is relatively more important than in cooler stars. The solid lines are simply approximate delimiters of the two ionization regions. The acoustic relevance of both regions is larger for hotter stars, as the distances $\omega^2-U^2$ are decreasing with the increase of temperature. Again, the acoustic relevance of region K2 is relatively increased for hotter stars. Right: the same differences $\omega^2 - U^2$ represented for the frequency $\nu_2$ ($\nu_2$ is represented in Figure \ref{fig4}). Here we have a similar picture to the  described above for the left panel. Because the turning point of the mode with $\nu_2$ is located deeper in the convective zone of the stars, these modes are relatively more affected by region K2 as we can see from the slopes of the "quasi-horizontal" fits. Finally, gray squares represent the differences $\omega^2 - U^2$ evaluated for the models of stars B and C evolved without microscopic diffusion and gray dashed lines the corresponding linear fits.}
\end{figure*}\label{fig5}

%\newpage

	\subsection{Exploring the phase shift of the reflected acoustic modes}

In this paper, we propose to use the phase shift, $\alpha(\omega)$;
\begin{equation} \label{eq5}
F({\omega}/L) = \pi \left( \frac{\alpha(\omega) + n}{\omega} \right) ,
\end{equation}
and its well-known relation with the structure of the envelope of solar-type stars \citep{1982Natur.300..242D,1989SvAL...15...27B,1989ASPRv...7....1V,1997ApJ...480..794L,1998MNRAS.295..344P,2001MNRAS.322..473L} to infer about the ionization processes in the outer layers of these stars. Equation \ref{eq5} is obtained when the internal asymptotic solutions that describe the oscillations are matched with the exact solutions on the surface of the star, where it is not possible to use asymptotic methods. Here, $F({\omega}/L)$ is determined by the profile of the sound speed over the stellar interior and it represents the time the acoustic wave takes to travel from the surface of the star to the turning point. $L=\ell+1/2$, with $n$ and $\ell$ as, respectively, the radial order and the degree of the mode. $\omega=2\pi\nu$ stands for the angular frequency of the mode, and $\nu$ is the cyclic frequency.  From Equation \ref{eq5}, it is possible to obtain for the derivative of the phase shift an expression, usually referred to as the seismic parameter $\beta$,

\begin{equation} \label{eq6}
\beta(\omega)=\beta(\nu) = - \nu^2 \frac{d}{d\nu} \left( \frac{\alpha}{\nu} \right) .
\end{equation}
This seismic parameter is known to be particularly sensitive to ionization regions and it can be written as \citep[e.g.][]{1987SvAL...13..179B}
\begin{equation} \label{eq7}
\beta(\nu) = \frac{\nu - n \left( \frac{ \partial \nu}{\partial n} \right) - L \left( \frac{\partial \nu}{\partial L} \right) }         {\frac{\partial \nu}{\partial n}}, 
\end{equation}
where is made explicit use of the oscillation mode frequencies. Therefore, using Equation \ref{eq7} it is possible to compute $\beta(\nu)$ from a table of frequencies by choosing an appropriate evaluation of the derivatives.

The effects of partial ionization on the oscillation spectrum, in the outer layers, can also be studied with the method of the acoustic potential. The complete set of differential equations describing the asymptotic theory of adiabatic stellar oscillations, in the context of the Cowling approximation, can be reduced to a Schr\"{o}dinger-type, time-independent equation \citep[e.g.,][]{1989ASPRv...7....1V, 2001MNRAS.322..473L}
\begin{equation} \label{eq8}
\frac{d^2 \psi}{d\tau^2} + (\omega^2 - U^2)\psi = 0 \, ,
\end{equation}
where $\tau$ is the acoustic depth (Equation \ref{eq4}) and $U^2$ is a function of the equilibrium model, usually known as the acoustic potential \citep[e.g.,][]{1995BCrAO..92...92B, 1997ApJ...480..794L, 2014ApJ...782...16B}. By solving Equation \ref{eq7}, we obtain a rigorous and robust representation of the reflection of acoustic waves in the upper layers of a solar-type star. This is similar to the well-known 1D standard potential barrier problem in quantum mechanics.
The derivative of the phase shift, the variable $\beta(\nu)$ (Equation \ref{eq6}), is tightly connected with the acoustic potential and thus can be related to the structure of the envelopes in solar-type stars. Therefore, it is a valuable tool for identifying the regions of partial ionization \citep[e.g.,][]{1987SvAL...13..179B, 1989SvAL...15...27B, 1994A&A...290..845L}.  Our aim is to use the acoustic potential and its relation with the function $\beta(\nu)$ to study the effects of ionization on the oscillation spectrum when considering different theoretical structures of the outer convective zone.

Figure \ref{fig4} shows the acoustic potential and the corresponding seismic parameters $\beta(\nu)$ computed for all the four benchmark theoretical models. In this figure, $\beta(\nu)$ was computed with two different methods: (1) from the structural variables of the envelopes of the theoretical models and (2) from a table of theoretical frequencies calculated with the ADIPLS code \citep{2008Ap&SS.316..113C}. The structural parameters of the stellar models are obtained in the Cowling approximation, i.e., without taking into account the effects of the gravitational potential. A consequence of the Cowling approximation is that the two theoretical seismic indicators, $\beta(\nu)$, exhibit a discrepancy which is well known also in the solar case \citep[e.g.][]{2001MNRAS.322..473L}. However, from Figure \ref{fig4}, it is seen that this gravitational potential effect influences the stars distinctively.  The horizontal dashed lines in Figure \ref{fig4} represent squared angular frequencies $\omega^2=(2\pi \nu)^2$. Regions where $U^2>(2\pi \nu)^2$ are propagation regions, whereas the values where $U^2=(2\pi \nu)^2$ represent the locations of the upper turning points of the acoustic modes. The vertical distance between the reflecting acoustic potential and the horizontal line defined by $\omega^2=(2\pi \nu)^2$ can be understood as a measure of the impact a given region inside the star has on the acoustic mode of oscillation with angular frequency $\omega$. The smaller this distance is, the larger the impact on the sound wave. The oscillatory feature in the seismic parameter $\beta(\nu)$ reflects in a nonlinear way the effects of the physical changes on the structure of the envelope produced on the oscillation spectrum. This quasi-periodic oscillatory feature is known to be tightly related to the characteristics of the ionization zones. As we can clearly see from Figure \ref{fig4}, its period and amplitudes change with the changes associated to the ionization regions K1 and K2.

\begin{figure*}
	\epsscale{1.2}
	\plotone{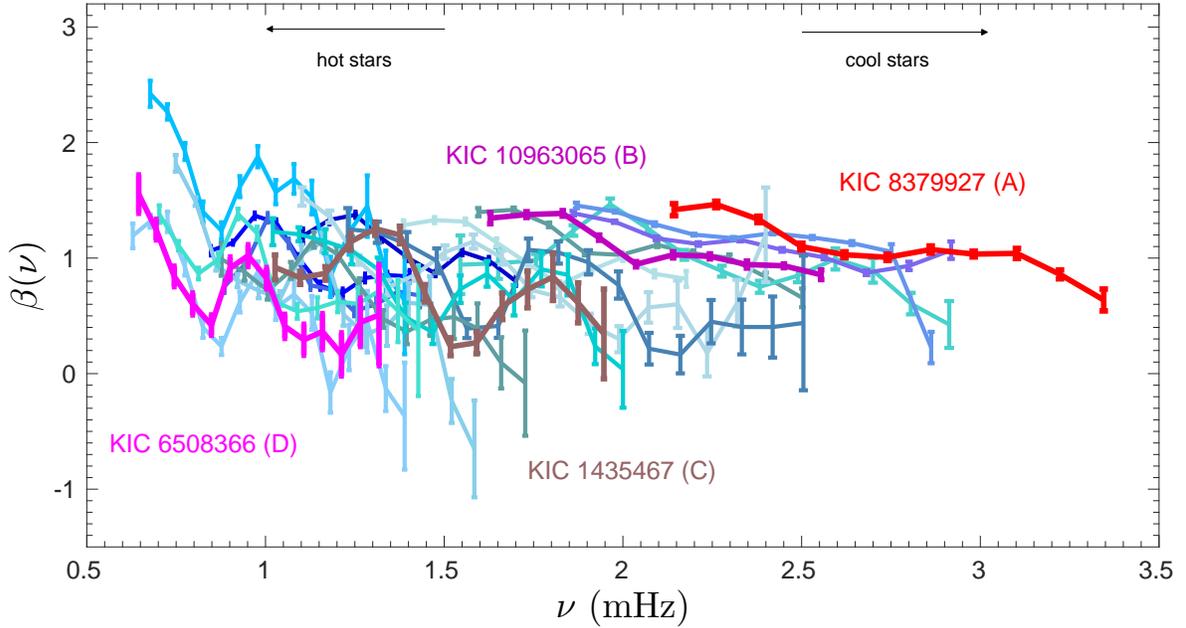}
	\caption{Observational signature $\beta(\nu)$ for 20 stars observed by {\it{Kepler}}. $\beta(\nu)$ is shown for modes with degree $l=1$. Observational oscillation mode frequencies are taken from \citet{2012A&A...543A..54A}. The four selected solar-type stars of this study are identified with the corresponding KIC values in the figure. The $\beta$ signatures of the hottest stars are located on the left-hand side of the figure, whereas $\beta$ signatures of the coolest stars are located on the right-hand side of the figure.}
\end{figure*}\label{fig6}

	\subsection{The Influence of the Ionization Regions in the Adiabatic Acoustic Oscillations}

Figure \ref{fig5} shows, for all the benchmark stars, a schematic representation of the distances $\omega^2-U^2$ for four different locations in the envelopes. These distances are indicated with dotted lines in Figure \ref{fig4}. The two panels of Figure \ref{fig5} refer to calculations made for two indicative values of the angular frequency: $\omega_1=2\pi\nu_1$, with $\nu_1 > \nu_{\text{max}}$ and $\omega_2=2\pi\nu_2$, with $\nu_2 < \nu_{\text{max}}$, with $\nu_1$ and $\nu_2$ being the references frequencies represented in Figure \ref{fig4}. Consequently, the acoustic mode characterized by the angular frequency $\omega_1$ has an upper turning point located in a more superficial part of the star than the acoustic mode with angular frequency $\omega_2$.
A careful inspection of Figure \ref{fig5} allows us two infer two important results from the theoretical data. The first result is expected and it supports the evidence that for all the studied stars, the partial ionization processes occurring in the region K1 will impact more the acoustic modes than partial ionization processes occurring in the region K2.  This is because the distance $\omega^2-U^2$ is smaller in the K1 region than in the K2 region. The second result is that with the increase of the effective temperature of the star, all ionization processes, affect more the acoustic modes (by an order of magnitude and within the $T_{\text{eff}}$ range of the benchmark stars). Furthermore, unlike what happens in cool stars, in hotter stars the influence of both K-shell related ionization regions turns out to have an impact of the same order on the acoustic oscillation modes.

As all the four stars have different values for the metallicity, we tested a potential effect of the metallicity on the acoustic properties of the theoretical models. Specifically, we computed the differences $\omega^2-U^2$ obtained from theoretical models without microscopic diffusion, thus keeping the metallicity (and helium) constant in the envelopes of the stars. The results of these tests without microscopic diffusion show that the increase of the impact of the K2 regions for hot stars cannot be attributed to a metallicity effect. To illustrate this statement, the distances $\omega^2-U^2$ were also represented without diffusion, in Figure \ref{fig5}, for the models of stars B and C, which are, respectively, the most and the less metallic stars.

\section{The Observable $\beta(\nu)$} \label{sec5}

The derivative of the phase shift represented by Equation \ref{eq7} leads to an interesting way of relating theory and observational data. In other words, this means that the function, $\beta(\nu)$, is also an observable \citep[e.g.][]{1989ASPRv...7....1V,1994A&A...290..845L, 1997ApJ...480..794L}. Figure \ref{fig6} shows the observational seismic parameter $\beta(\nu)$ for the 20 \textit{Kepler} stars of Figure \ref{fig1}. The sinusoidal component of the observed $\beta(\nu)$, known to be tightly connected with the partial ionization processes in the Sun and solar-type stars, reflects the increase of the magnitude of the ionization with the mass and radius of the star, given that these are main-sequence stars (see Figure \ref{fig2}). This trend manifests itself in an exacerbation of the prominent quasi-periodic variation of $\beta(\nu)$ with the increase of the star's effective temperature. For a set of 12 stars, we compared the theoretical seismic signature $\beta(\nu)$ with the observational $\beta(\nu)$, computed directly from tables of frequencies taken from \citet{2012A&A...543A..54A}. The results of this comparison are shown in Figure \ref{fig7}.
Naturally, some stars exhibit a better agreement between the theoretical models and the observations. Stars with masses closer to values of the Sun tend to show better agreements between theory and observations. Because the derivative of the phase shift is particularly sensitive to ionization processes in the outer layers of these stars, this could mean that the accuracy of the equation of state is not enough to describe solar-type stars with masses slightly different from the Sun. In particular, the effects of heavy elements ionization may be underestimated in the current equations of state.   In a recent result from \citet{2017arXiv170804937B}, a new equation of state, the SAHA-S equation of state \citep{2013CoPP...53..392G}, has been tested in the case of a solar model. The differences in the profile of the first adiabatic exponent $\Gamma_1$ for the ionization region K2 (i.e., in the temperature range  from $\log$T $\sim 5.6$ to $\log$T $\sim 6.1$) are of the order of $10^{-3}$. It is expected that for hotter stars, a similar comparison will achieve even larger differences.

Other potential sources of disagreement between the theoretical and the observational $\beta(\nu)$ may be related to radiative effects. Heavy elements abundances are important to determine the opacities of stars. Opacities, in  turn, also impact the oscillation spectrum of solar-type stars as it is well known, for instance, from the solar abundance problem \citep{2008PhR...457..217B, 2009ApJ...705L.123S} where the neon abundance seems to play an important role \citep{2005Natur.436..525D}. Selective atomic diffusion processes, such as gravitational settling of chemical elements, also have impact in the pulsation spectrum of stars \citep{2000A&A...360..603T, 2004MNRAS.350.1022P, 2009A&A...506...69A}.  Table \ref{table:3} shows a comparison of the initial metallicities with the surface metallicities of the models. Also shown in Table \ref{table:3} is the initial helium abundance and the final helium abundances of the models at the surface. Results from helioseismology \citep[e.g.][]{1998ApJ...504..539T} allow us to assume that in cool solar-type stars the effects of gravitational settling of chemical elements can be considered generally small. However, with the increase of the mass of the star, i.e., for hotter stars, selective diffusion effects such as radiative accelerations of heavy elements may become important.

\begin{figure*}
	\centering
	\includegraphics[width=0.3\textwidth]{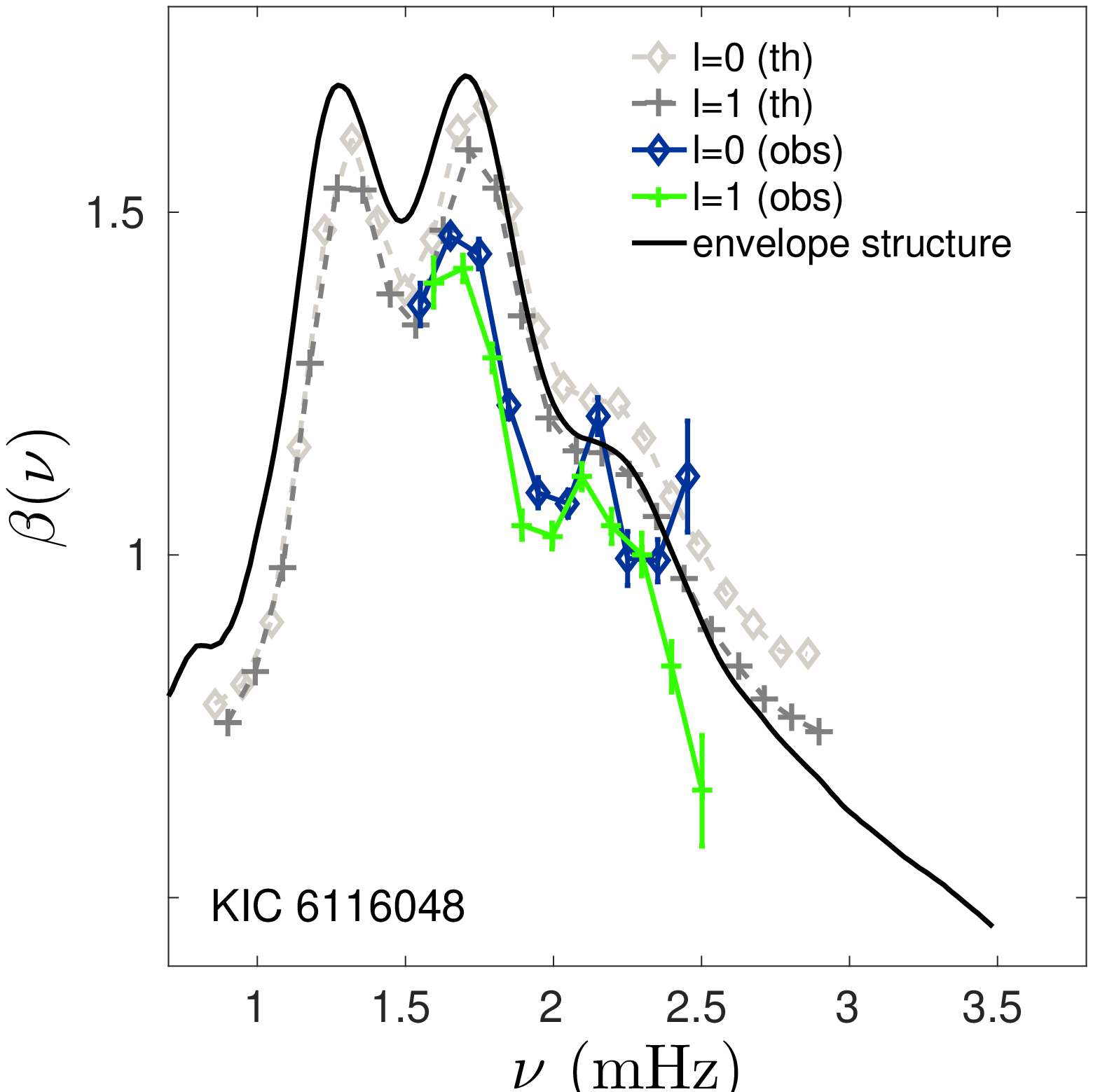}
	\includegraphics[width=0.3\textwidth]{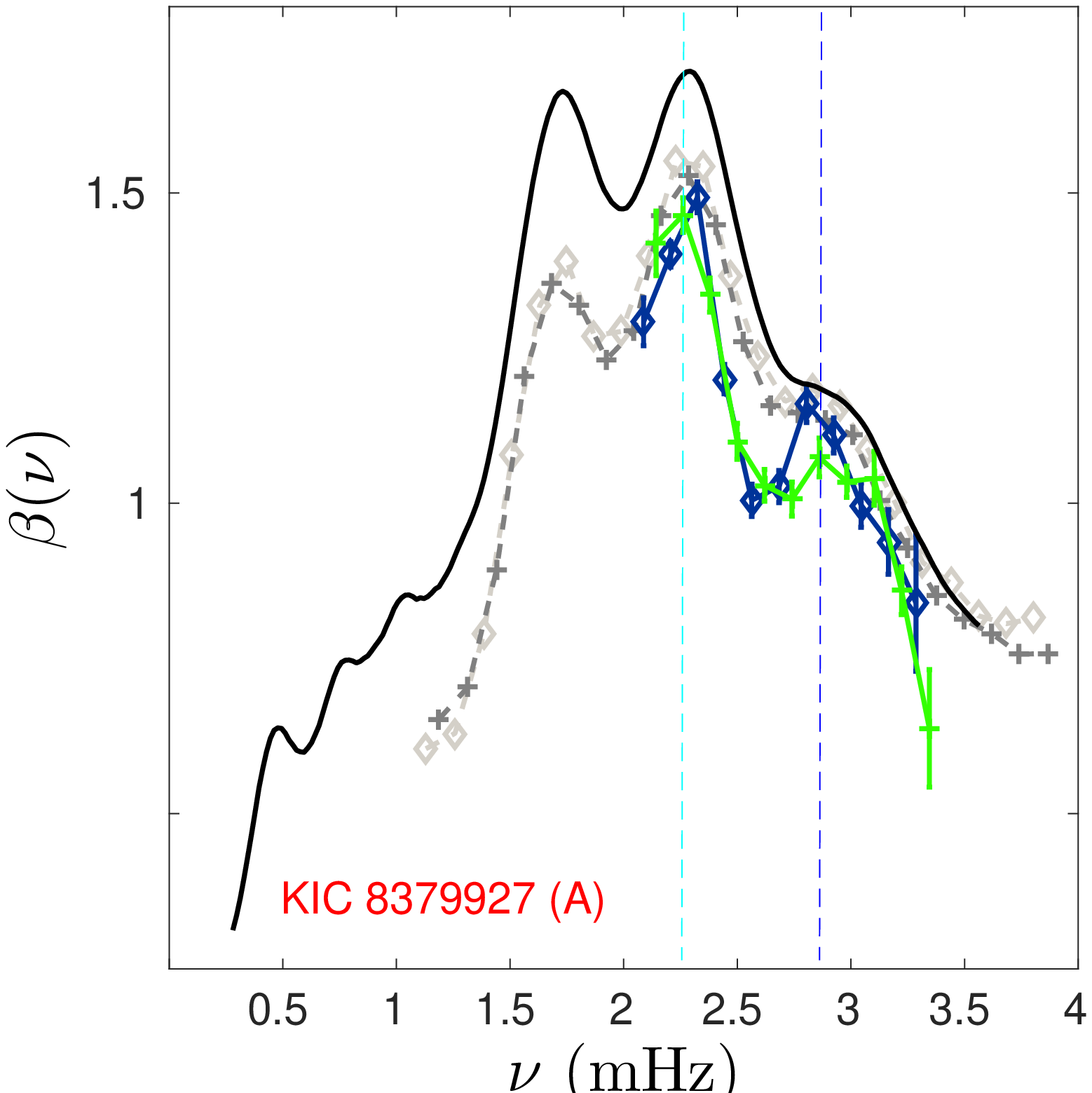}
	\includegraphics[width=0.3\textwidth]{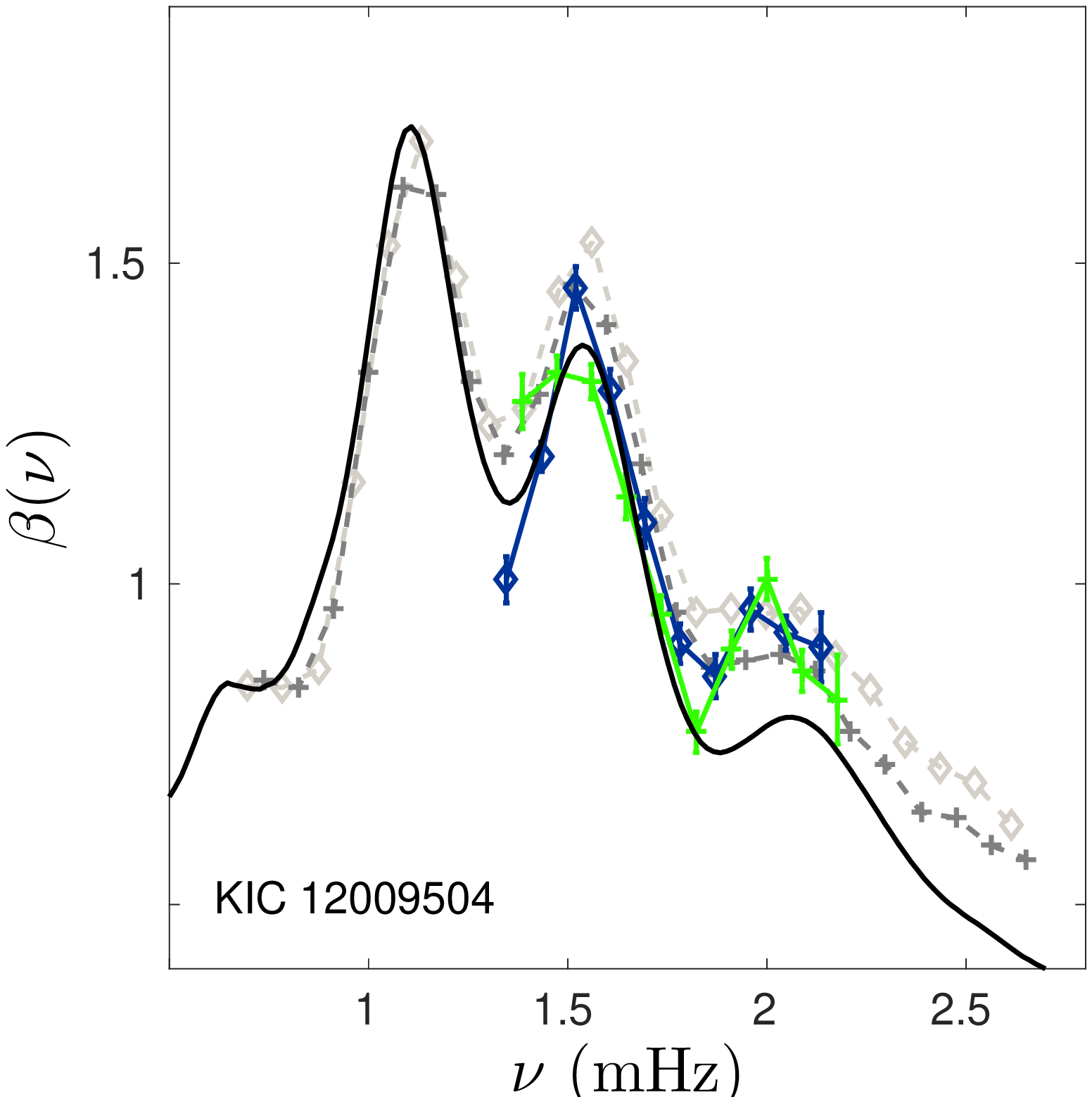}\\
	\includegraphics[width=0.3\textwidth]{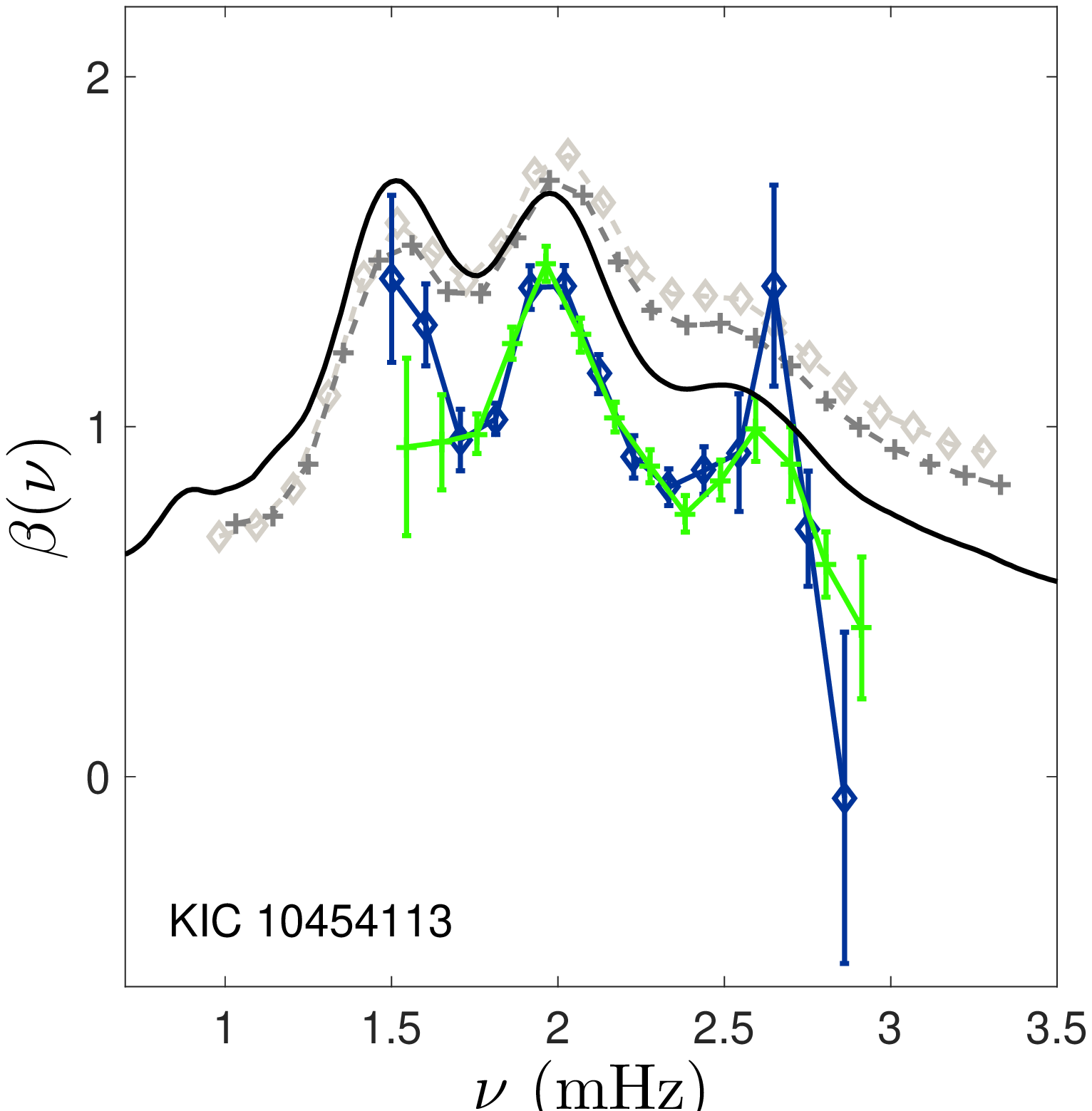}
	\includegraphics[width=0.3\textwidth]{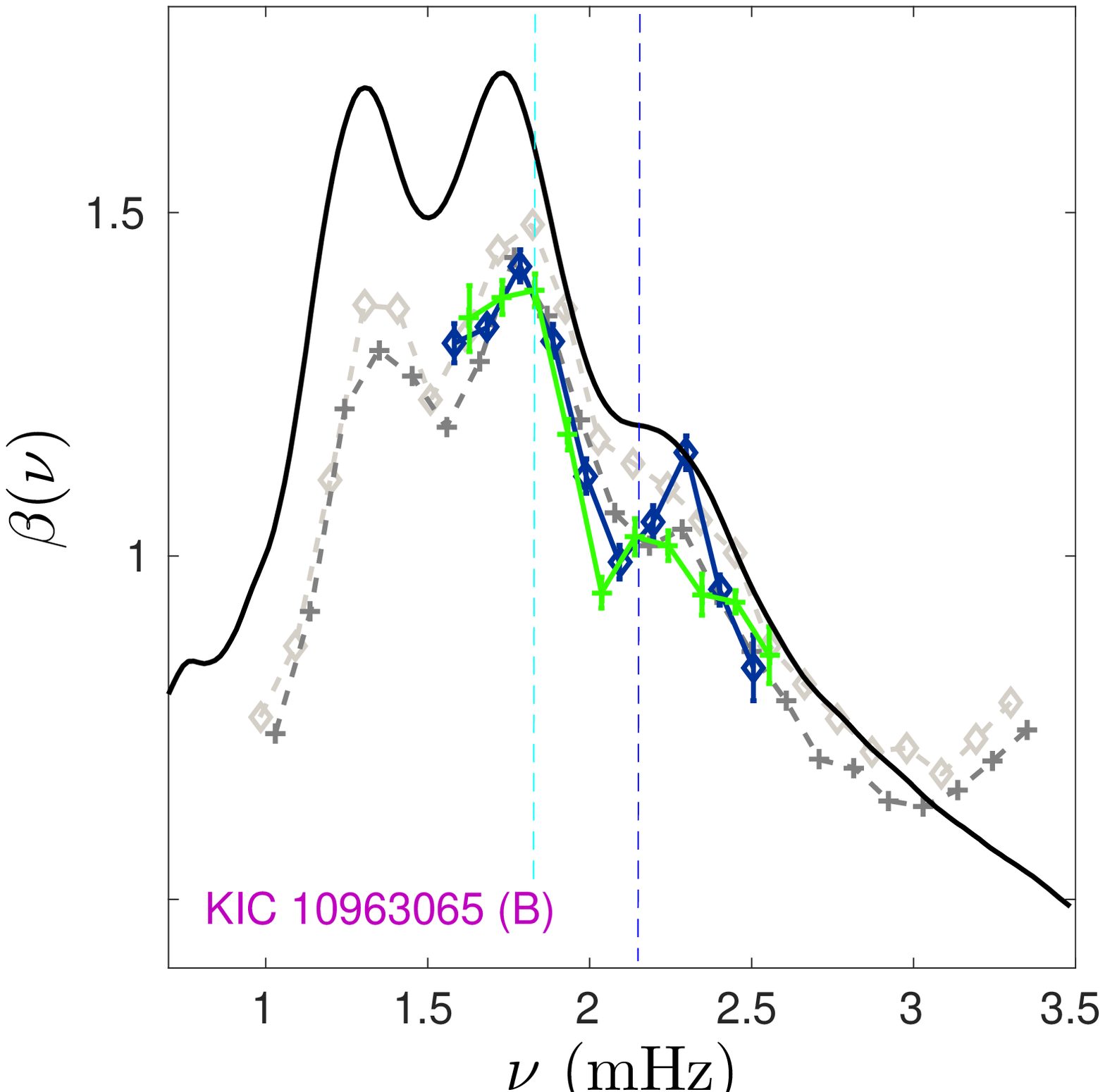}
	\includegraphics[width=0.3\textwidth]{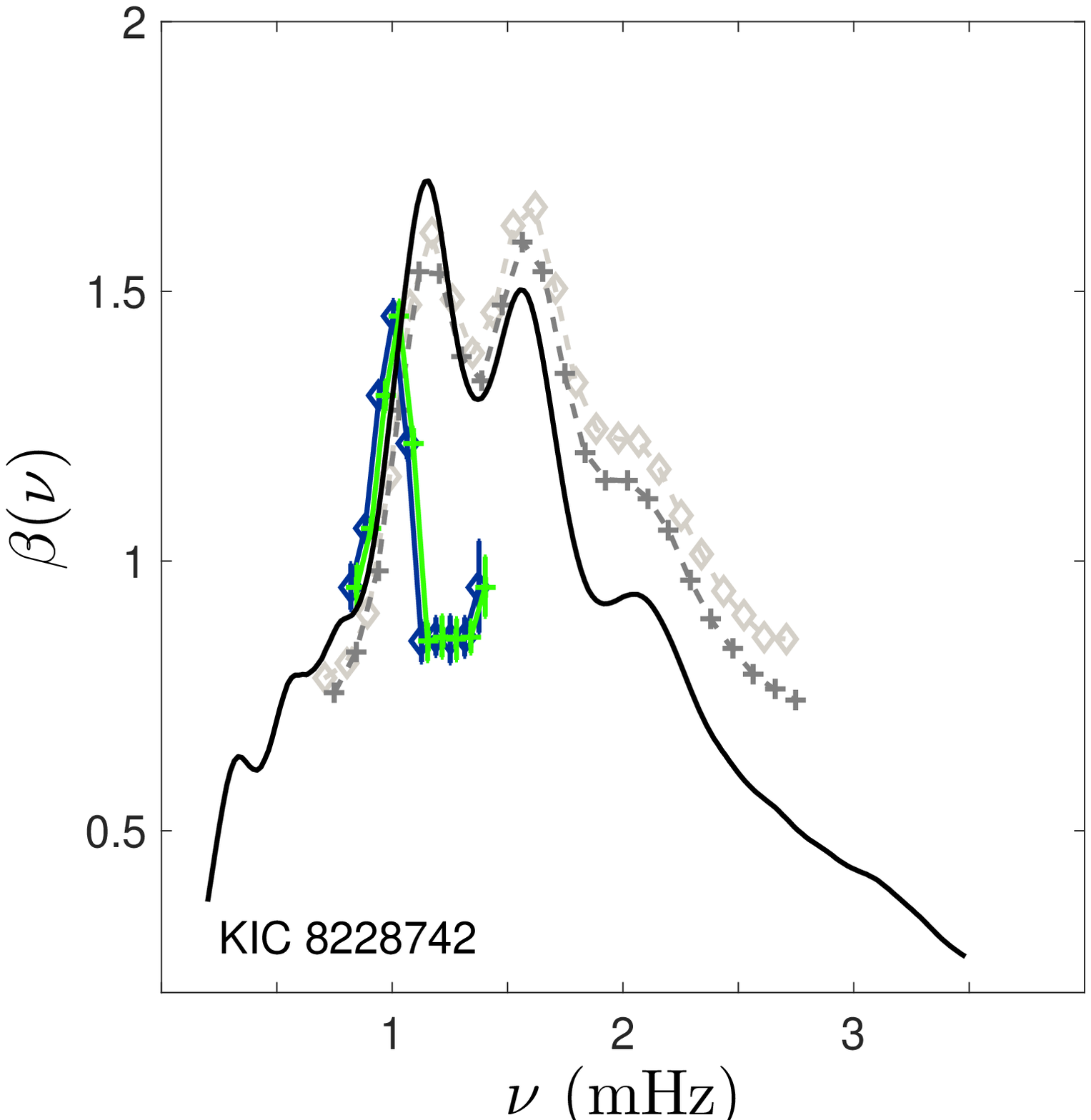}\\
	\includegraphics[width=0.3\textwidth]{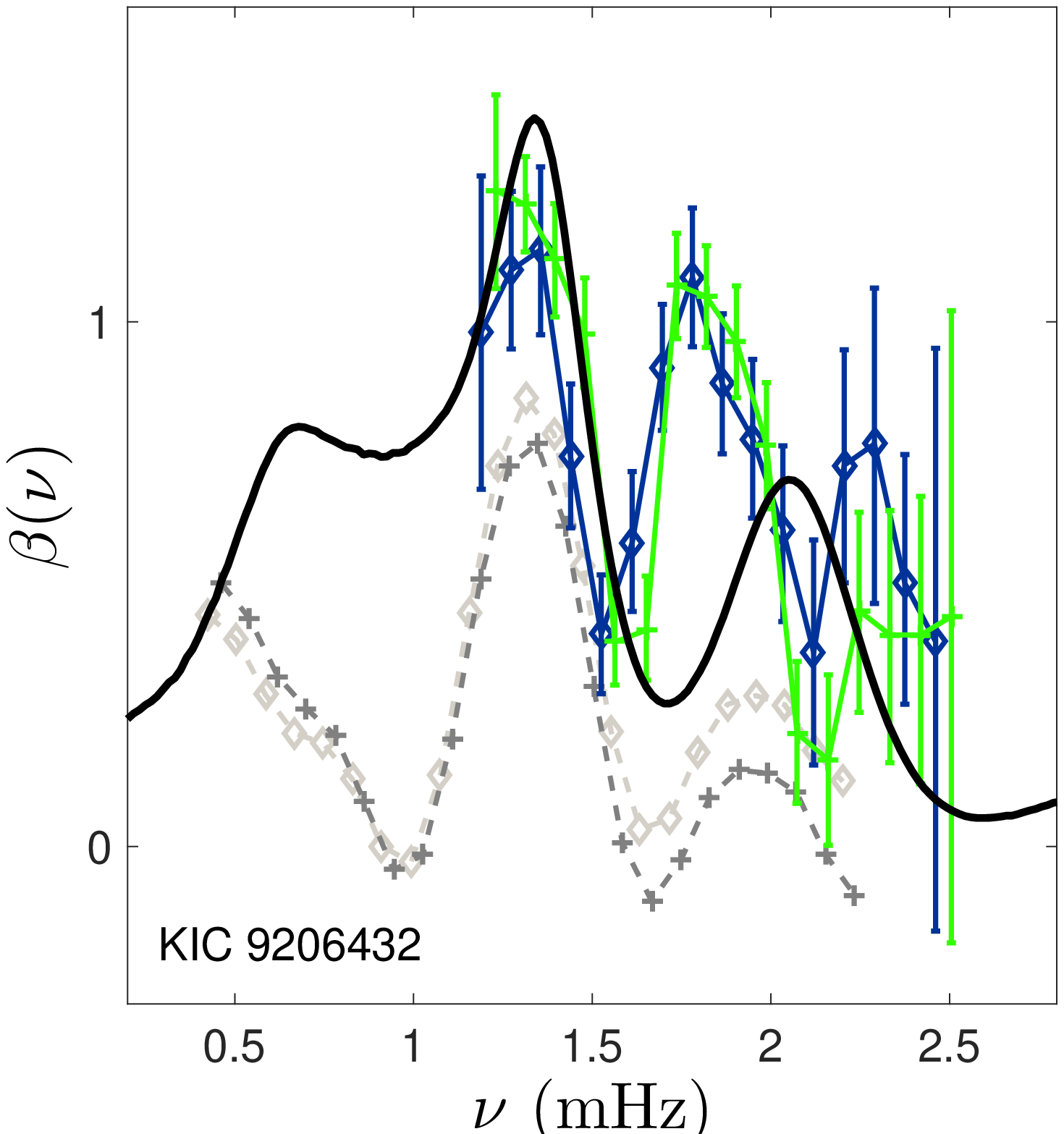}
	\includegraphics[width=0.3\textwidth]{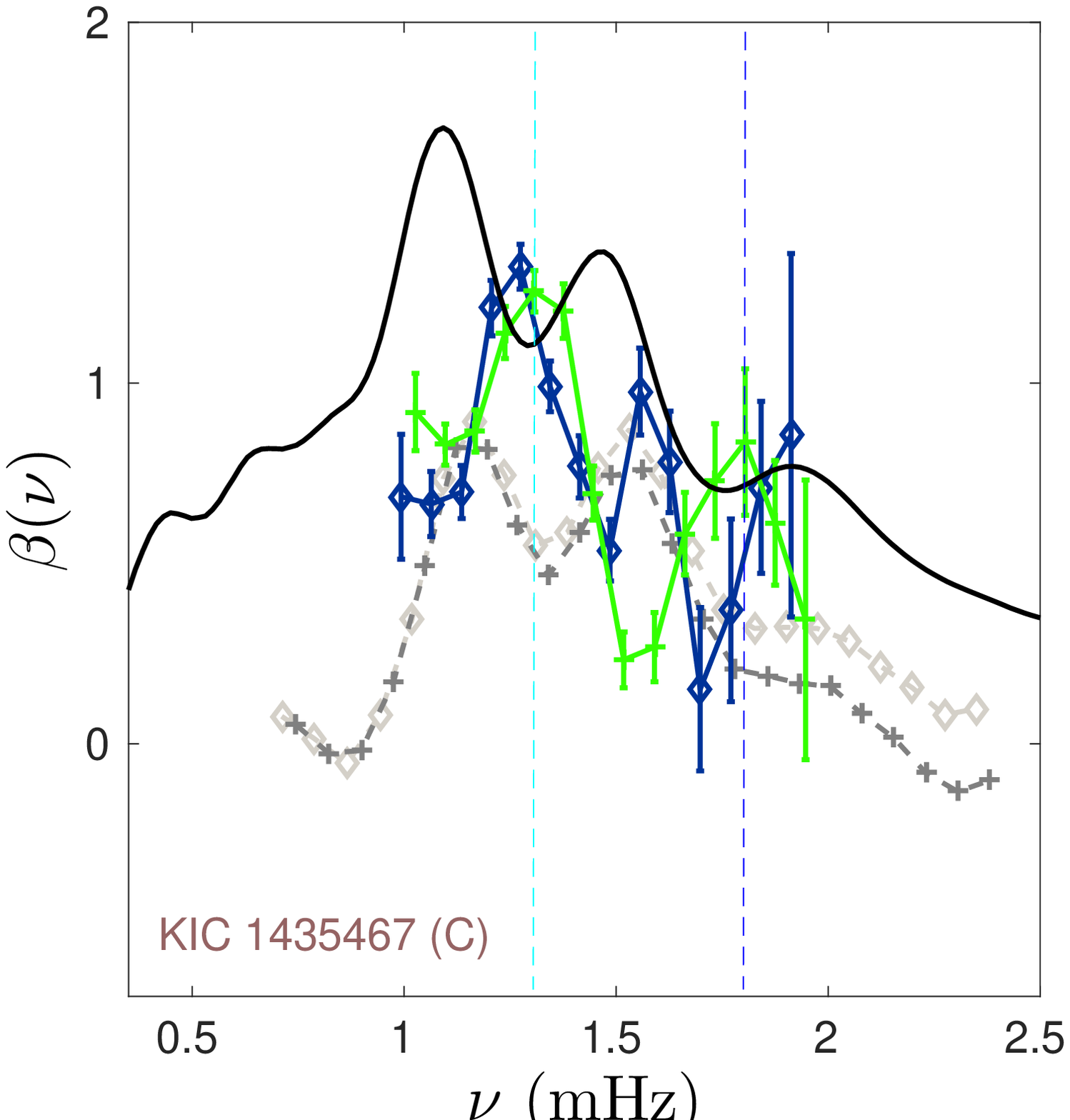}
	\includegraphics[width=0.3\textwidth]{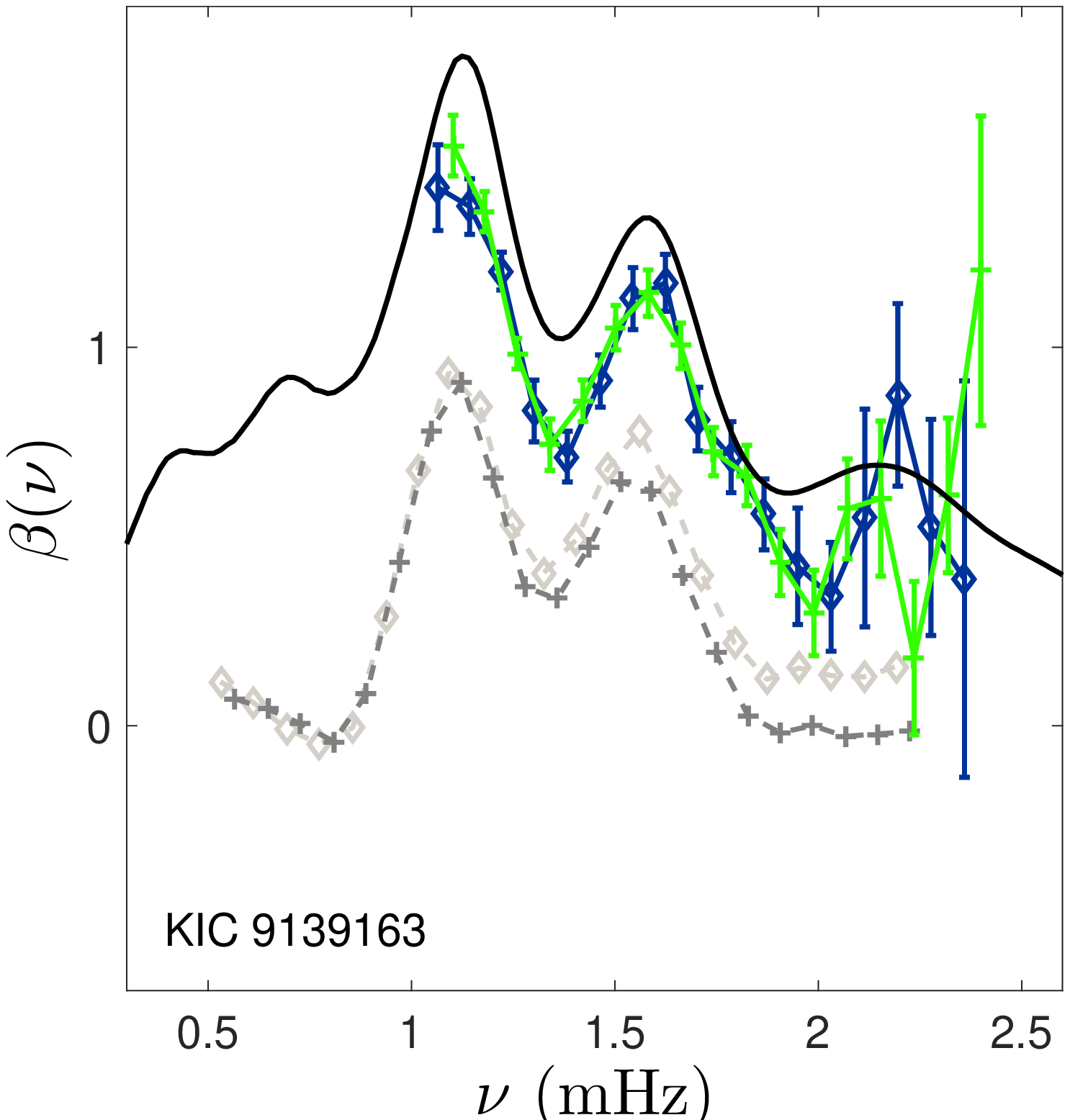}\\
	\includegraphics[width=0.3\textwidth]{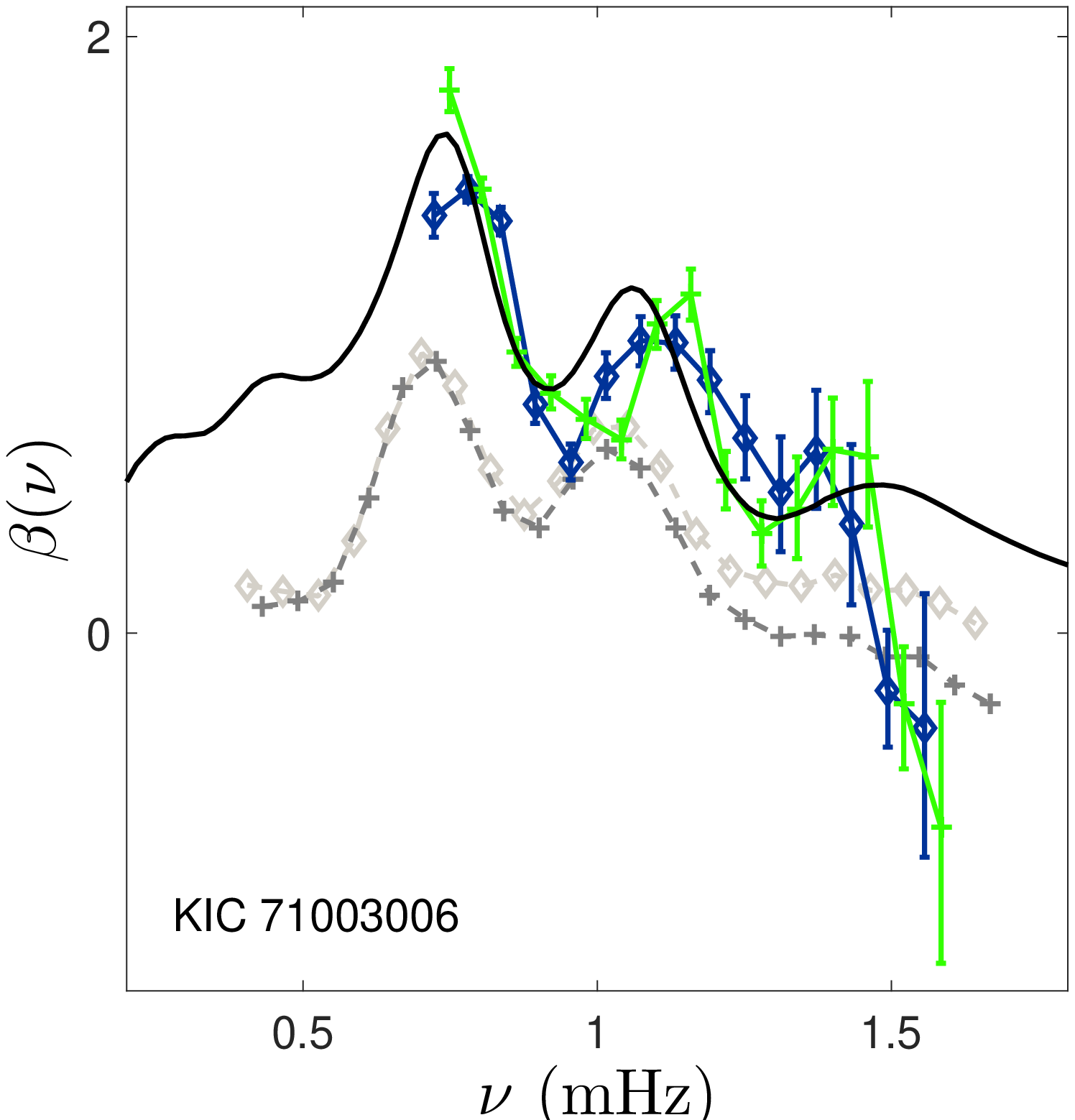}
	\includegraphics[width=0.3\textwidth]{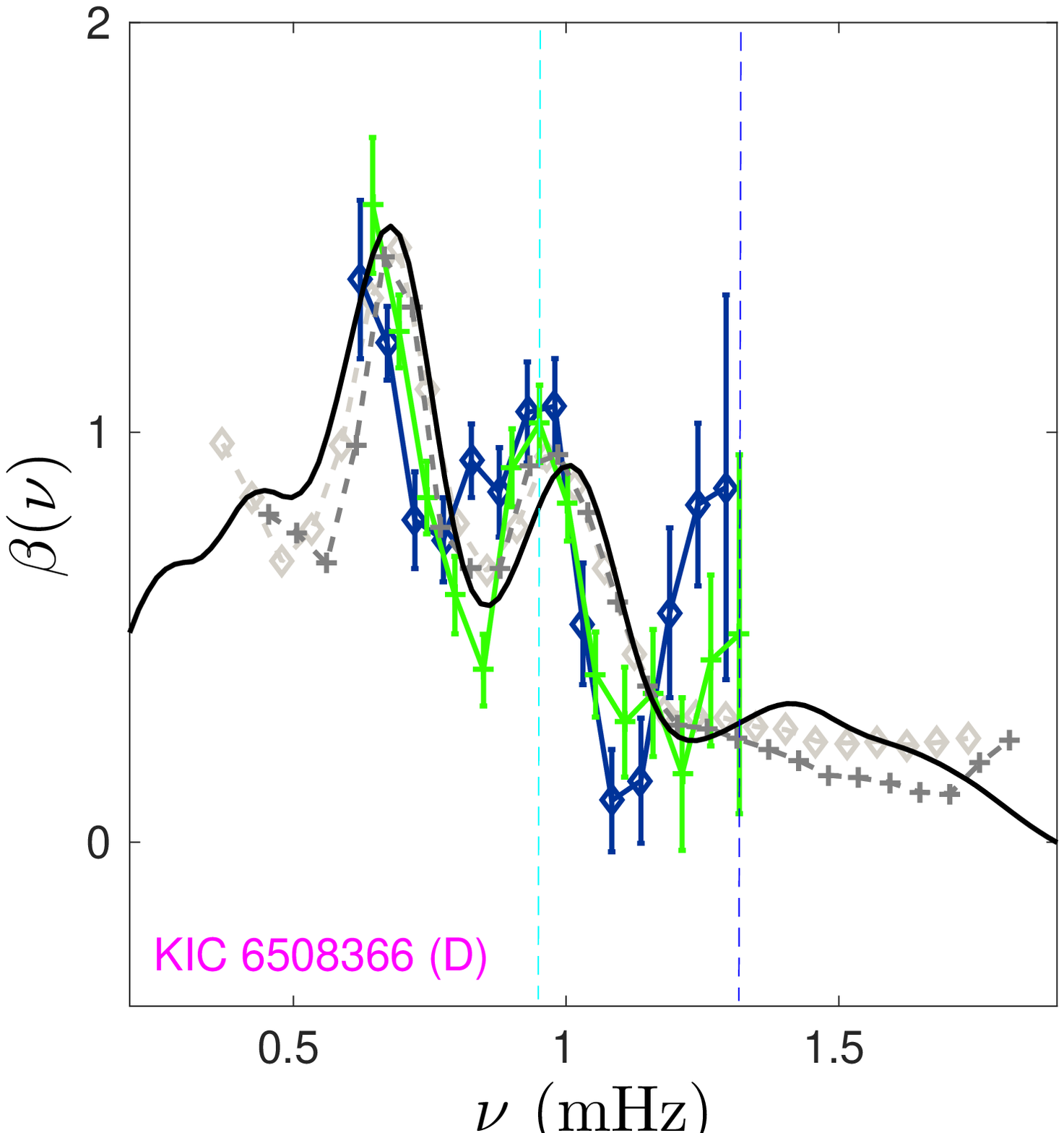}
	\includegraphics[width=0.3\textwidth]{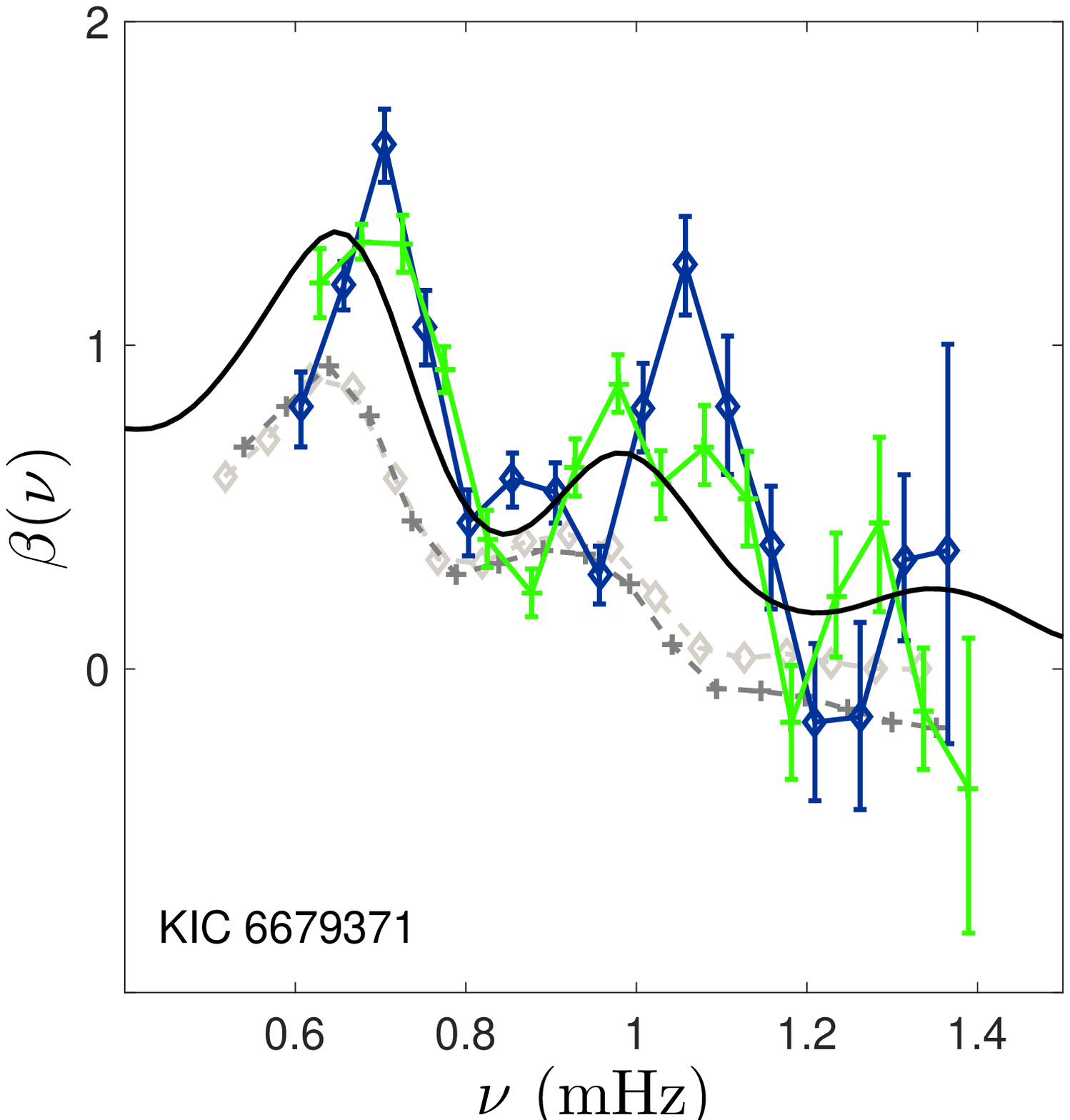}
	\caption{Comparison between the theoretical seismic parameter $\beta(\nu)$ and the seismic observable $\beta (\nu)$. Solid black lines represent the parameter $\beta(\nu)$ computed for the envelope structure of the theoretical models. Gray lines show $\beta(\nu)$ evaluated from theoretical tables of frequencies. Finally, the observational signatures are indicated  by dark blue and green colors. The four benchmark stars of this study are shown in the middle panel. Vertical dashed lines signalize the locations of the reference frequencies, $\nu_1$ and $\nu_2$, discussed in Section \ref{sec4}.}
\end{figure*}\label{fig7}

The results of Section \ref{sec4}, specifically the fact that in hotter stars the influence of both K-shell related ionization regions could be comparable, might imply some important consequences in the study of the seismically acoustic sharp features in hot F-type stars. A variation of the sound speed can be considered seismically sharp if its radial extension is less or approximately equal to the wavelength of the acoustic mode.  It is known that rapid variations of the sound speed lead to an oscillatory signature in the frequencies. These oscillatory features also manifest themselves in different seismic parameters such as, for example, the second differences $\delta_2 \nu_{n,\ell}=\nu_{n+1, \ell} -2 \nu_{n,\ell} + \nu_{n-1, \ell}$ \citep[e.g.,][]{1990LNP...367..283G} or in the seismic parameter $\beta(\nu)$ \citep[e.g.,][]{1994MNRAS.268..880R, 1997ApJ...480..794L}. From the periodicity analysis of the quasi-periodic oscillation in the seismic parameter $\beta(\nu)$ it is possible to  relate its period with the corresponding scattering region \citep{2014ApJ...782...16B, 2017MNRAS.466.2123B}. The adequacy of such method rely, among other factors, on the ability to decontaminate the quasi-periodic signal from contributions or interferences of other relevant physical processes occurring in the same region. Table \ref{table:4} shows the results of the sinusoidal fits to the oscillatory feature of the observational seismic parameter $\beta(\nu)$ for modes with degree $\ell=1$. The choice of $\ell=1$ modes is justified by the fact that these are the modes with the highest S/N. The sinusoidal fits were obtained with the same technique used by \citet{2017MNRAS.466.2123B}.
When comparing the locations obtained for the acoustic depths in Table \ref{table:4} with the corresponding locations in the acoustic potentials of each star, we verify that for cooler stars the fitted quasi-periodic oscillation of $\beta(\nu)$ relates to a location that clearly coincides with the ionization region K1 (usually known as the region of the second ionization of helium). This is expected because as we described above, in cooler stars, the impact of region K1 dominates over the impact of region K2. As the effective temperature of the star increases, the quasi-periodic oscillation in $\beta(\nu)$ undergoes changes and these changes seem to be related to the growing influence of the region K2.  The locations obtained with the sinusoidal fits in the hottest stars are moving toward the interior of the star, because possibly they are influenced by the two partial ionization regions, K1 and K2 (see Figure \ref{fig4}).

A theoretical analysis based on the acoustic potentials of the stars is a powerful and robust complementary method to investigate the so called acoustic glitches in solar-type stars. Particularly, it can help us with the interpretation of acoustic glitches in hot stars. In these hot stars, the oscillatory signal usually attributed to the helium ionization \citep[e.g.,][]{2017ApJ...837...47V} can be contaminated with effects of the partial ionization of heavy elements.

Finally, it is well known that the theoretical adiabatic oscillation mode frequencies are usually overestimated for high values of the radial order \citep[e.g.][]{2012MNRAS.422L..43G}. It is possible to introduce descriptive corrections to the frequencies \citep[e.g.][]{2008ApJ...683L.175K}; however, as these corrections may somehow mask or disguise the phenomena addressed in this study, we opted by not including them.

%%%%%%%%%%%%%%%%%%%%%%%%%%%%%%%%%%%%%%%%%%%%%%%%%%%%%%%%%%%%%%%%%%%%%%%%%%%%%%%%%%%%%%%%
%%%%%  TABLE 4
%%%%%%%%%%%%%%%%%%%%%%%%%%%%%%%%%%%%%%%%%%%%%%%%%%%%%%%%%%%%%%%%%%%%%%%%%%%%%%%%%%%%%%%%
\begin{deluxetable}{c|cc} 
	\tablecaption{Results of the sinusoidal fits to the observable $\beta(\nu)$ \label{table:4}}
	\tablecolumns{3}
	\tablenum{4}
	\tablewidth{0pt}
	\tablehead{
		\colhead{Star Id.} &
		\colhead{$\tau$ ($\ell =1$)} & 
		\colhead{ $\tau_{\text{ He, obs}}$ \tablenotemark{(a)}} 
	}
	\startdata 
	A &   596 s    &   $703^{+21}_{-22}$ s      \\
	B &   815 s    &   $793^{+47}_{-40}$ s      \\
	C &   1025 s   &   $1112^{+37}_{-35}$ s      \\
	D &   1498 s   &   $1527^{+61}_{-58}$ s      \\
	\enddata
	\tablenotetext{}{(a) \citet{2017ApJ...837...47V}}
\end{deluxetable}
%%%%%%%%%%%%%%%%%%%%%%%%%%%%%%%%%%%%%%%%%%%%%%%%%%%%%%%%%%%%%%%%%%%%%%%%%%%%%%%%%%%%%%%%
%%%%%  TABLE 4 - END
%%%%%%%%%%%%%%%%%%%%%%%%%%%%%%%%%%%%%%%%%%%%%%%%%%%%%%%%%%%%%%%%%%%%%%%%%%%%%%%%%%%%%%%%

\section{Conclusions} \label{sec6}

In this work, we perform a study of the partial ionization processes occurring in the convective envelopes of four stars. These four stars were chosen among the {\it{Kepler}} targets with the highest quality asteroseismic data. They are benchmark stars in the sense that they cover all the observed range of the average large-frequency separation for the {\it{Kepler}} dwarfs. 

Our study is based on a combination of different theoretical methods and simultaneously makes use of an observable seismic parameter to scrutinize the theoretical results. We analyze the reflecting acoustic potentials together with the seismic parameter $\beta(\nu)$ because it is well known that physical changes in the structure of the envelope are reflected in the characteristics of the seismic parameter $\beta(\nu)$, which, in turn, is a proxy of the phase shift of the reflected acoustic waves. In addition, we provide a detailed analysis of the partial ionization profiles for seven chemical elements (hydrogen, helium, carbon, nitrogen, oxygen, neon, and iron). To derive the relevant ionization regions, we use the gradient of the mean effective ionic charge and we found two relevant ionization regions, both linked to the ionization of the electrons of the atomic K-shell. One of these regions, the closest to the surface of the star is related to the K-shell ionization of helium. The second relevant region, located deeper in the convective zone, seems to be related to K-shell ionizations of period 2 elements which in this specific study, are carbon, nitrogen, oxygen, and neon. Moreover, we found that in cooler stars the impact of the ionization region related to the heavy elements on the frequencies is small, but in hotter stars the impact of this region on the mode frequencies can be comparable to the impact of the region related to the second ionization of helium. 
F-stars are known to display different patterns of magnetic activity \citep[e.g.,][]{2014A&A...562A.124M}. Magnetic activity and  rotation are apparently deeply linked to the physics of the convective zones. Detailed studies about the microphysics of stellar interiors can help us to understand and describe the ever-improving observational data sets.

Finally, we want to highlight the interest of a combined study like ours, if a few more observed mode frequencies in the low-frequency part of the spectrum become available. It is precisely these low frequencies that are needed to probe this interesting zone related to the K-shell ionizations of heavy elements in the convective envelopes of solar-type stars.
\vspace{1mm}

\acknowledgments

\vspace{5mm}
We want to thank to the anonymous referee the many useful suggestions, comments and remarks that led to an improvement of the quality of the manuscript.
We are also grateful to P. Morel for making available the CESAM code for stellar evolution and 
to Jordi Casanellas for the modified version of the same code and for his valuable help. Acknowledgments also to  J. Christensen-Dalsgaard for his Aarhus adiabatic pulsation code (ADIPLS). 
The authors thank the Funda\c c\~ao para a Ci\^encia e Tecnologia (FCT), Portugal, for the financial support to the  Center for Astrophysics and Gravitation (CENTRA),  Instituto Superior T\'ecnico,  Universidade de Lisboa,  through the grant No. UID/FIS/00099/2013.
This work was also supported by grants from "Funda\c c\~ao para a Ci\^encia e Tecnologia" (SFRH/BD/74463/2010).
%\software{CESAM \citep{1997A&AS..124..597M}, ADIPLS \citep{2008Ap&SS.316..113C}}

%\begin{thebibliography}{}
%
%\bibitem[Astropy Collaboration et al.(2013)]{2013A&A...558A..33A} Astropy Collaboration, Robitaille, T.~P., Tollerud, E.~J., et al.\ 2013, \aap, 558, A33 
%\bibitem[Bertin \& Arnouts(1996)]{1996A&AS..117..393B} Bertin, E., \& Arnouts, S.\ 1996, \aaps, 117, 393 
%\bibitem[Corrales(2015)]{2015ApJ...805...23C} Corrales, L.\ 2015, \apj, 805, 23
%\bibitem[Ferland et al.(2013)]{2013RMxAA..49..137F} Ferland, G.~J., Porter, R.~L., van Hoof, P.~A.~M., et al.\ 2013, \rmxaa, 49, 137
%\bibitem[Hanisch \& Biemesderfer(1989)]{1989BAAS...21..780H} Hanisch, R.~J., \& Biemesderfer, C.~D.\ 1989, \baas, 21, 780 
%\bibitem[Lamport(1994)]{lamport94} Lamport, L. 1994, LaTeX: A Document Preparation System, 2nd Edition (Boston, Addison-Wesley Professional)
%\bibitem[Schwarz et al.(2011)]{2011ApJS..197...31S} Schwarz, G.~J., Ness, J.-U., Osborne, J.~P., et al.\ 2011, \apjs, 197, 31  
%\bibitem[Vogt et al.(2014)]{2014ApJ...793..127V} Vogt, F.~P.~A., Dopita, M.~A., Kewley, L.~J., et al.\ 2014, \apj, 793, 127  
%
%\end{thebibliography}

\bibliography{bib_pap4}

\end{document}